\documentclass[10pt,conference,letterpaper]{IEEEtran}
\usepackage{amsthm}
\usepackage{amsmath,bm}
\usepackage{amssymb}
\usepackage{graphicx}
\usepackage{latexsym}
\usepackage{makecell}
\usepackage{array}
\usepackage{float}
\usepackage{epstopdf}
\usepackage[linesnumbered,ruled,vlined]{algorithm2e}
\usepackage{multirow}
\usepackage[justification=centering]{caption}
\usepackage{color}
\usepackage{algorithmicx}
\usepackage{algpseudocode}
\usepackage{amsmath}
\usepackage[vlined,ruled]{algorithm2e}
\usepackage{xcolor}
\usepackage{enumerate}
\usepackage{comment}

\usepackage{url}

\newcommand{\tabincell}[2]{\begin{tabular}{@{}#1@{}}#2\end{tabular}}

\newtheorem{myDef}{Definition}

\title{Updates-Aware Graph Pattern based Node Matching}
\author{

{Guohao Sun{\small$^{1}$}, Guanfeng Liu{\small$^{1}$}, Yan Wang{\small$^{1}$} and Xiaofang Zhou$^{2}$}%
\vspace{1.6mm}\\
\fontsize{10}{10}\selectfont\itshape
$^{1}$\,Department of Computing, Macquarie University, Sydney, NSW 2109, Australia\\
$^{2}$\,School of Information Technology and Electrical Engineering, The University of Queensland, Brisbane 4072, Australia\\
\fontsize{9}{9}\selectfont\ttfamily\upshape
$^{1}$\,guohao.sun@students.mq.edu.au; $^{1}$\,\{guangfeng.liu, yan.wang\}@mq.edu.au;  $^{2}$\,zxf@itee.uq.edu.au%
\vspace{1.2mm}\\
\fontsize{10}{10}\selectfont\rmfamily\itshape
}

\begin{document}
\maketitle
\begin{abstract}
\emph{Graph Pattern based Node Matching} (GPNM) is to find all the matches of the nodes in a data graph $G_D$ based on a given pattern graph $G_P$. GPNM has become increasingly important in many applications, e.g., group finding and expert recommendation. In real scenarios, both $G_P$ and $G_D$ are updated frequently. However, the existing GPNM methods either need to perform a new GPNM procedure from scratch to deliver the node matching results based on the updated $G_P$ and $G_D$ or incrementally perform the GPNM procedure for each of the updates, leading to low efficiency. Therefore, there is a pressing need for a new method to efficiently deliver the node matching results on the updated graphs. In this paper, we first analyze and detect the elimination relationships between the updates. Then, we construct an Elimination Hierarchy Tree (EH-Tree) to index these elimination relationships. In order to speed up the GPNM process, we propose a graph partition method and then propose a new updates-aware GPNM method, called UA-GPNM, considering the single-graph elimination relationships among the updates in a single graph of $G_P$ or $G_D$, and also the cross-graph elimination relationships between the updates in $G_P$ and the updates in $G_D$. UA-GPNM first delivers the GPNM result of an initial query, and then delivers the GPNM result of a subsequent query, based on the initial GPNM result and the multiple updates that occur between two queries. The experimental results on five real-world social graphs demonstrate that our proposed UA-GPNM is much more efficient than the state-of-the-art GPNM methods.
\end{abstract}

%
\section{Introduction}
\subsection{Background}
Graph Pattern Matching (GPM) is to find all the matching subgraphs of a pattern graph $G_P$ in a data graph $G_D$. In order to address the low-efficiency issue in the conventional NP-Complete GPM methods \cite{ullmann1976algorithm, cordella2001improved, garey2002computers}, Fan et al., proposed \emph{Bounded Graph Simulation} (BGS) \cite{fan2010graph}, which has fewer restrictions but more capacity to efficiently extract more useful subgraphs because it supports simulation relations instead of an exact match of edges and nodes. In BGS, each node in $G_D$ and $G_P$ has a label (e.g., representing a person's job title), and each edge in $G_P$ is labeled with either a positive integer $k$ or a symbol ``*". $k$ is the constraint of the maximal shortest path length of a match in $G_D$ and ``*" indicates that there are no path length constraints. Then, the match of an edge could be a path if the start node and the end node of the path in the data graph have the same labels as the corresponding nodes of the edge in the pattern graph respectively. In social networks, on average, any two people can be connected in about six hops \cite{milgram1967small}. Therefore, $k$ is usually set as a small integer in social networks \cite{fan2010graph}.

The GPM methods discussed above aim to find the entire subgraphs in $G_D$. However, in some applications, such as group finding \cite{lappas2009finding} and expert recommendation \cite{morris2010people, brynielsson2010detecting}, people are willing to find a group of nodes based on a specified structure between them, leading to the \emph{Graph Pattern based Node Matching} (GPNM) problem \cite{liu2007identifying}, with an example discussed below. 

\begin{figure}
\setlength{\abovecaptionskip}{0.2cm}
\setlength{\belowcaptionskip}{0.2cm}
\centering
\includegraphics[width=2.8in,height=1.6in]{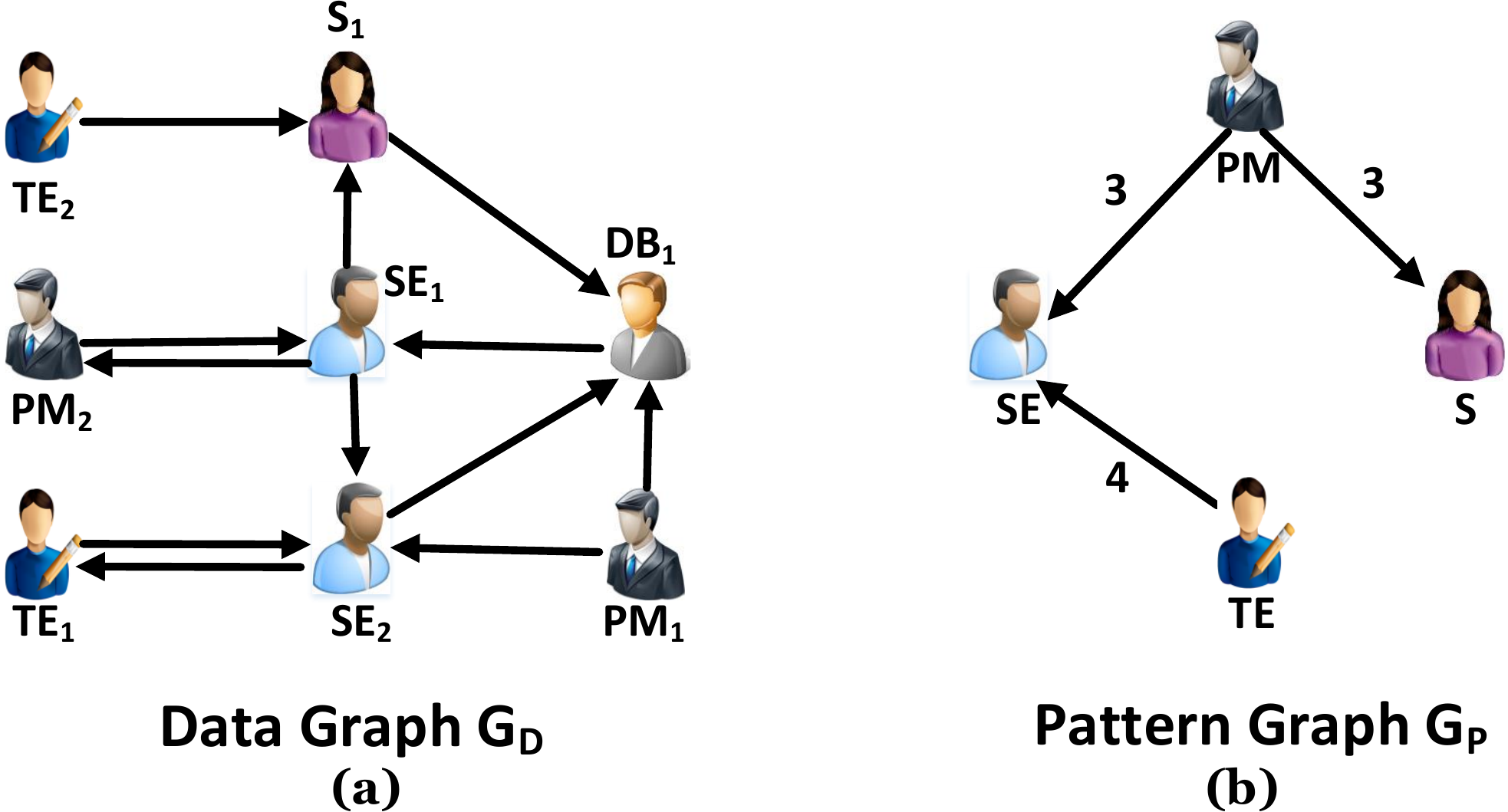}
\caption{Graph Pattern based Node Matching}
\label{fig:pnmatch}
\vspace{-0.15in}
\end{figure}

\noindent \textbf{Example 1 (GPNM Problem):} Fig. \ref{fig:pnmatch}(a) depicts a data graph $G_D$, where each node denotes a person, labeled with his job title, e.g., \emph{Project Manager} ($PM$), \emph{Database Developer} ($DB$), \emph{Software Engineer} ($SE$), \emph{Test Engineer} ($TE$), or \emph{Secretary} ($S$). Each edge indicates a collaboration relationship. A pattern graph $G_P$ is given in Fig. \ref{fig:pnmatch}(b), where an IT project needs four types of people, namely, $PM$, $SE$, $TE$, and $S$ respectively. In BGS \cite{fan2010graph}, the integer on an edge shows the constraint of the maximum path length between two nodes. For example, in Fig. \ref{fig:pnmatch}(b), a $PM$ needs to connect with an $SE$ and an $S$ within 3 hops respectively. The GPNM results are shown in TABLE \ref{tab:GPMResult}.

\begin{table}[h]
\tiny
\setlength{\abovecaptionskip}{0.2cm}
\setlength{\belowcaptionskip}{0.2cm}
\centering
\vspace{-0.1in}
\caption{The node matching results of Example 1}
\begin{tabular}{|c|c|} \hline
Nodes in $G_P$&Matching nodes in $G_D$\\ \hline
$PM$ & $PM_1$\\ \hline
$SE$ & $SE_1$, $SE_2$\\ \hline
$S$ & $S_1$ \\ \hline
$TE$ & $TE_1$, $TE_2$\\
\hline\end{tabular}
\label{tab:GPMResult}
\vspace{-0.1in}
\end{table}

The existing GPM methods can be applied to solve the GPNM problem. However, they need to deliver the entire matching subgraphs, rather than the matching nodes only, which incurs a high time complexity \cite{fan2010graph, fan2011adding}. Therefore, Fan et al., \cite{fan2013diversified} proposed a method to find matching nodes only based on a given pattern graph. Although their method can reduce query processing time, it does not consider the updates of $G_P$ and $G_D$ that commonly exist in real scenarios \cite{berger2006framework}. Even if there exists only one update in a small size pattern graph, for this update in pattern graph, the existing GPNM methods have to recompute the matching results in the data graph starting from scratch, which leads to much query processing time. In addition, the query structures of the pattern graphs given by billions of users in social networks are changed very frequently. Therefore, the updates to patterns are in high frequency. Therefore, it is necessary and significant to consider the updates in pattern graph and data graph to improve the efficiency of node matching. For example, in group finding in Online Social Networks (OSNs) \cite{lappas2009finding}, the joining of new users or the withdrawal of existing users in OSNs results in the updates of $G_D$. When facing each of such updates, the existing GPNM methods \cite{liu2007identifying, fan2013diversified} have to perform a new GPNM procedure from scratch, leading to low efficiency.

In order to improve efficiency, the state-of-the-art GPNM methods, called INC-GPNM \cite{Sun2018incremental} and EH-GPNM \cite{sun2019incremental}, have been proposed. INC-GPNM first incrementally records the shortest path length range between different types of labels in $G_D$ and then identifies the affected area of $G_D$ w.r.t. the updates of $G_P$ and $G_D$. Thus, INC-GPNM can improve the efficiency of GPNM when $G_P$ and $G_D$ are updated. However, in a large-scale social graph that is updated with a high frequency, INC-GPNM is still computationally expensive as it ignores the relationships that exist among the updates in both $G_P$ and $G_D$, and thus, when facing any update, it has to perform an incremental GPNM procedure for each of the updates. EH-GPNM considers the updates in $G_D$ only. When facing updates in the pattern graph, it still has to perform the incremental GPNM procedure for each of the updates in the pattern graph. Therefore, a new efficient GPNM method is in demand.

\subsection{Motivations and Problems}\label{sec:motivations}
In real scenarios, nodes and edges in both $G_P$ and $G_D$ are usually frequently updated over time. For example, in the application of group finding in social graphs, different queries can have different constraints and/or structures, which leads to the updates of $G_P$, and the joining of new users in OSNs leads to the updates of $G_D$. However, not all the updates in a pattern graph $G_P$ or a data graph $G_D$ essentially affect the GPNM matching results. Below Example 2 illustrates the details of our motivations.

\begin{figure}
\setlength{\abovecaptionskip}{0.cm}
\setlength{\belowcaptionskip}{-0.cm}
\centering
\includegraphics[width=2.8in,height=2.9in]{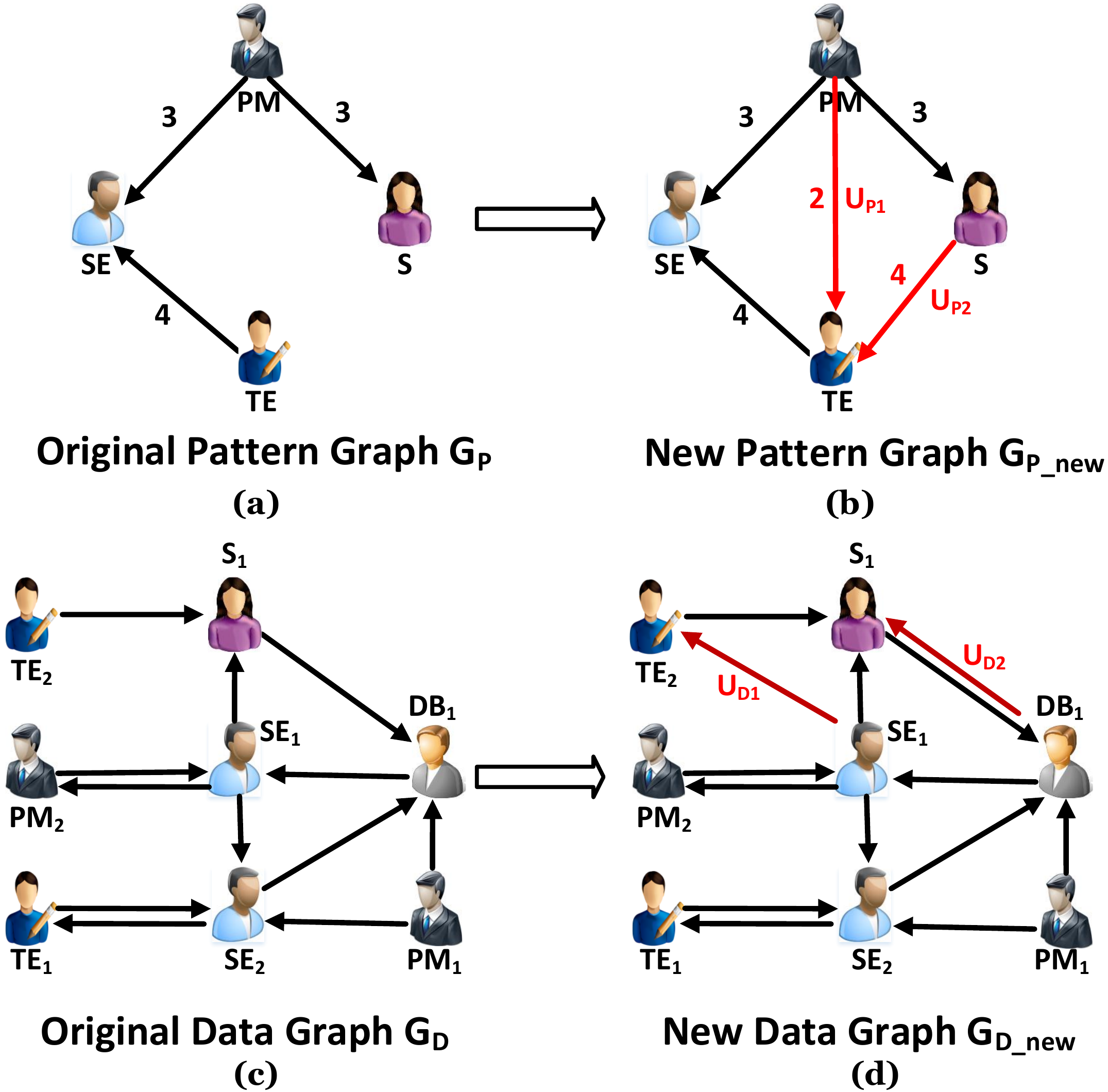}
\caption{Updates-aware GPNM}
\label{fig:incmatch}
\vspace{-0.15in}
\end{figure}

\noindent \textbf{Example 2 (Updates-aware GPNM):} Based on the pattern graph and data graph shown in Fig. \ref{fig:incmatch}(c) and Fig. \ref{fig:incmatch}(a) respectively, the original GPNM matching results are shown in Table \ref{tab:GPMResult}. Suppose there are two updates in the pattern graph, where $PM$ needs to be associated with a $TE$ within 2 hops (denoted as $U_{P1}$ in Fig. \ref{fig:incmatch}(b)), and an $S$ needs to be associated with a $TE$ within 4 hops (denoted as $U_{P2}$ in Fig. \ref{fig:incmatch}(b)). And there are also two updates in data graph, where $SE_1$ establishes the collaboration relationship with $TE_2$ (denoted as $U_{D1}$ in Fig. \ref{fig:incmatch}(d)) and $DB_1$ establishes the collaboration relationship with $S_1$ (denoted as $U_{D2}$ in Fig. \ref{fig:incmatch}(d)). The new pattern graph $G_{P\_new}$ and new data graph $G_{D\_new}$ are shown in Fig. \ref{fig:incmatch}(b) and Fig. \ref{fig:incmatch}(d) respectively.

Based on these two updated graphs, the state-of-the-art incremental GPNM method \cite{Sun2018incremental} has to apply the incremental procedure four times because there are a total of four updates in $G_P$ and $G_D$, leading to low efficiency. However, in practice, one update can be eliminated by another update. It is easy to understand that in each single graph, if one edge (node) is firstly removed from (or inserted into) $G_D$ ($G_P$) and then inserted back to (or removed from) $G_D$ ($G_P$), the effects of the two updates can eliminate each other. Therefore, there may exist elimination relationships among the updates in a single graph of $G_P$ or $G_D$, and we term this kind of elimination relationships of a single graph as \emph{single-graph elimination relationships}. More importantly, one update in a graph may eliminate an update in another graph, we term this kind of elimination relationships as \emph{cross-graph elimination relationships}. In Example 2, although in update $U_{P1}$, a $PM$ needs to be associated with a $TE$ within 2 hops, it indeed leads to no change in the GPNM results. This is because in another update $U_{D1}$, $SE_1$ happens to establish the collaboration with $TE_2$, making all the $PMs$ in the data graph be connected with a $TE$ within 2 hops. Therefore, the effects of $U_{P1}$ and $U_{D1}$ eliminate each other.

This example motivates us to develop a new GPNM solution which considers the elimination relationships among the updates to efficiently answer GPNM queries. When facing an updated pattern graph and an updated data graph, we can compute the GPNM result for the original pattern graph, and then deliver the new GPNM result by analyzing the elimination relationships among the updates, instead of performing the incremental GPNM procedure for each of the updates.

Such a new GPNM solution is significant for the social graph searches in large-scale and frequently updated social networks, such as Facebook and Twitter. For example, on Facebook, on average, within each minute, 400 new users join in, 510,000 comments are posted, 317,000 statuses are updated, and 147,000 photos are uploaded\footnote{https://sproutsocial.com/insights/facebook-stats-for-marketers/}. 

In this new solution, there are three major challenges. Firstly, it is non-trivial to identify the elimination relationships among the updates because there exist both single-graph elimination relationships and cross-graph elimination relationships. Therefore, the first challenge of our work is \textbf{CH1}: \emph{how to effectively detect the elimination relationships of the updates.} Secondly, if update $U_a$ eliminates update $U_b$, and update $U_b$ eliminates update $U_c$, there exists a hierarchical structure of them, which applies to all the elimination relationships. As it is computationally expensive to deliver GPNM results by investigating each of the elimination relationships among the updates, it is beneficial to build up an index to record the hierarchical structure of all the elimination relationships. Therefore, the second challenge of our work is \textbf{CH2}: \emph{how to build up an index structure to record the hierarchical structure of all the elimination relationships covering both single-graph elimination relationships and cross-graph elimination relationships, which supports the development of an efficient algorithm to deliver the GPNM results by making use of the index.} Thirdly, in the GPNM procedure, we need to compute the shortest path length between any two nodes, which is very time-consuming. Therefore, the third challenge of our work is \textbf{CH3}: \emph{how to efficiently compute the shortest path length between any two nodes to speed up the GPNM procedure.}

\subsection{Contributions}
In this paper, we propose an efficient GPNM method to answer GPNM queries with multiple updates in both pattern graph and data graph. To the best of our knowledge, our method is the first GPNM solution that considers both the single-graph elimination relationships and cross-graph elimination relationships. The characteristics and contributions of our work are summarized as follows.

(1) Targeting \textbf{CH1}, we propose effective methods to detect the single-graph elimination relationships and cross-graph elimination relationships.

(2) Targeting \textbf{CH2}, we build up an Elimination Hierarchy Tree (EH-Tree) to index the hierarchical structure of all the different types of elimination relationships, which helps enhance query processing efficiency.

(3) Targeting \textbf{CH3}, we propose a graph partition method and based on the method we further propose a more efficient Updates-Aware GPNM algorithm called UA-GPNM.

(4) The experiments conducted on five real-world social graphs demonstrate that our UA-GPNM with graph partition strategy significantly outperforms the state-of-the-art GPNM methods \cite{Sun2018incremental, sun2019incremental} by reducing the the query processing time with an average of \emph{58.60\%} and \emph{35.29\%} respectively.

The rest of this paper is organized as follows. We first review the related work in Section \ref{sec:relatedwork}. Then we introduce the necessary concepts and formulate the main problem in Section \ref{sec:problemdefinition}. Section \ref{sec:Elimination Relationship} analyzes the elimination relationships. Section \ref{sec:partition} introduces the partition method. Section \ref{sec:UA-GPNM} proposes the new algorithm, UA-GPNM. Section \ref{sec:experiment} discusses the experimental results, and Section \ref{sec:conclusion} concludes the paper.

\section{Related Work}\label{sec:relatedwork}
The existing methods can be classified into two categories based on their delivered matching results: i.e., (1) \emph{Graph Pattern Matching (GPM)}, and (2) \emph{Graph Pattern based Node Matching (GPNM)}. In this section, we review these two categories respectively.

\noindent \textbf{GPM:} GPM is to find all the matching subgraphs of $G_P$ in $G_D$. For example, the algorithm in \cite{ullmann1976algorithm} is the most famous method for the subgraph isomorphism. In the light of the intractability of the NP-complete problem of subgraph isomorphism, an approximate solution BGS \cite{fan2010graph} has been studied to find inexact matching subgraphs. In the application of community finding, Fang et al., \cite{fang2016effective} proposed a method which aims to return an attributed community for an attributed graph, in which the attributed community is a subgraph which satisfies both structure cohesiveness and keyword cohesiveness. Fang et al., \cite{lai2017scalable} studied scalable subgraph enumeration in MapReduce, considering that existing solutions for subgraph enumeration are not sufficiently scalable to handle large graphs. 

However, social graphs are frequently updated \cite{berger2006framework}, and it is computationally expensive to perform a new procedure from scratch to find matching subgraphs when facing any updates. Therefore, Fan et al., \cite{fan2013incremental} proposed an incremental approximate method to find the matching subgraphs. The complexity of this method is more accurately characterized in terms of the size of the area affected by the updates of data graphs, rather than the size of the entire input. Song et al., \cite{song2014event} propose a new notion, “event pattern matching” on dynamic graphs. They study the semantics and efficient online algorithms for the event pattern matching. In the application of cyber security, Choudhury et al., \cite{choudhury2015selectivity} present a new subgraph isomorphism algorithm in streaming graphs. They regard cyber attacks as a subgraph pattern, and apply the subgraph distributional statistics collected from the streaming graph to determine the query processing strategy. Semertzidis et al., \cite{semertzidis2016durable} focused on labeled graphs that evolve over time. They find the matches that exist for the longest period of time. Sun et al., \cite{sun2017mining} extended incremental methods to find maximal cliques that contain vertices incident to an edge which has been inserted. Fan et al., \cite{fan2017incremental} further proposed incremental algorithms for four types of typical pattern graphs, which can reduce the computations on big graphs and minimize unnecessary re-computation. Ma et al., \cite{ma2017fast} proposed a method to find dense subgraphs in temporal networks. They focueds on a special class of temporal networks, where the weights associated with edges regularly vary with timestamps. Li et al., \cite{li2018persistent} aimed to identify the communities that are persistent over time in a temporal network, in which every edge is associated with a timestamp. In addition, Li et al., \cite{li2018efficient} proposed a method to seek cohesive subgraphs in a signed network, in which each edge can be positive or negative, denoting friendship or conflict respectively. Li et al., \cite{li2019time} proposed a solution to efficiently answer subgraph search in streaming graph data. In the method, they designed concurrency management strategies to improve system throughput. Das et al., \cite{das2019incremental} proposed change-sensitive algorithms to maintain the set of subgraphs in dynamic graphs. They showed nearly tight bounds for the magnitude of change in the set of subgraphs and the time complexity of enumerating the change is proportional to the magnitude of the change. Dias et al., \cite{dias2019fractal} proposed Fractal, a high performance and high productivity system for supporting distributed GPM applications. Fractal employs a dynamic (auto-tuned) load-balancing based on a hierarchical and locality-aware work stealing mechanism, allowing the system to adapt to different workload characteristics.\\
\noindent \textbf{GPNM:} Applying the existing GPM methods to solve the GPNM problem incurs a high time complexity as they need to deliver the entire matching subgraphs in $G_D$ \cite{fan2010graph, fan2011adding}. Therefore, several GPNM methods have been proposed, which aim to find some nodes based on a specified structure between those nodes, such as group finding \cite{lappas2009finding} and expert recommendation \cite{morris2010people}. Some of them \cite{liu2007identifying}, \cite{zou2009distance}, \cite{marian2005adaptive} are proposed to find matches of a specific node via \emph{subgraph isomorphism}, which has the exponential complexity. To improve efficiency, Tong et al., \cite{tong2007fast} proposed a "Seed-Finder" method that identifies approximate matches for certain pattern nodes. This method only requires cubic time. Based on BGS, Fan et al., \cite{fan2013diversified} revised graph patterns to support a specific output node and define functions to measure match relevance and diversity. Motivated by network analysis applications, Fan et al., \cite{fan2016adding} proposed quantified matching for a specific pattern node, in which they extend traditional graph patterns with counting quantifiers. 

To address the GPNM problem when graphs are updated over time, INC-CPNM and EH-GPNM have been proposed in \cite{Sun2018incremental} and \cite{sun2019incremental}. INC-GPNM first builds an index to incrementally record the shortest path length range between different label types in $G_D$, and then identifies the affected nodes of $G_D$ in GPNM w.r.t. the updates of $G_P$ and $G_D$. Moreover, based on the proposed index structure and novel search strategies, INC-GPNM can efficiently deliver node matching results taking the updates of $G_P$ and $G_D$ as input, and can greatly reduce the query processing time. EH-GPNM \cite{sun2019incremental} realized there may exist single-graph elimination relationships in the data graph. It can deliver the GPNM results without performing the incremental procedure for each of the updates in the data graph.\\
\noindent \textbf{Summary:} The existing methods in the above two categories face the efficiency issue when answering GPNM queries with the updates in both pattern graphs and data graphs. Firstly, the GPM methods cannot be applied in GPNM because of the low efficiency of delivering the entire subgraph structures. Secondly, the state-of-the-art GPNM methods INC-GPNM \cite{Sun2018incremental} and EH-GPNM \cite{sun2019incremental} cannot offer good efficiency either. INC-GPNM has to perform the incremental procedure for each of the updates, which is still computationally expensive in a large-scale graph that is updated frequently. Although EH-GPNM realized there may exist single-graph elimination relationships in data graph, it ignores the single-graph elimination relationships in pattern graph and the cross-graph elimination relationships. When facing any update in the pattern graph, EH-GPNM still has to perform the incremental GPNM procedure for each of the updates in the pattern graph. 

\section{Preliminaries}\label{sec:problemdefinition}

In this section, we introduce the concepts of data graph and pattern graph, and the problem of GPNM and the problem of Updates-Aware GPNM. Table \ref{tab:notations} lists the notations used in this paper.
\begin{table}[t]
\vspace{-0.1in}
\tiny
\setlength{\abovecaptionskip}{0.2cm}
\centering
\caption{Notations used in this paper}
\begin{tabular}{|l|l|} \hline
\textbf{Notation}&\textbf{Meaning}\\ \hline
$G_D$/$G_P$& a data graph/pattern graph\\ \hline
$G_{D\_new}$/$G_{P\_new}$ & an updated data/pattern graph \\ \hline
$\bigtriangleup{G_D}$/$\bigtriangleup{G_P}$ & the updates of $G_D$/$G_P$ \\ \hline
$e(v_i, v_j)$ & a directed edge from $v_i$ to $v_j$ \\ \hline
$V$/$E$ & a set of vertices/edges in $G_D$ \\ \hline
$V_P$/$E_P$& a set of vertices/edges in $G_P$ \\ \hline
$f_e(u, v)$ & the bounded path length on $e(u, v)$ in $G_P$\\ \hline
$M(G_P, G_D)$ & the matching result of $G_P$ in $G_D$ based on BGS\\ \hline
$IQuery$/$SQuery$ & the GPNM result of the initial/subsequent query\\ \hline
$\bigtriangleup{G_{D_E}^{+}}$ / $\bigtriangleup{G_{D_E}^{-}}$& the insertions / deletions of edges for $G_D$\\ \hline
$\bigtriangleup{G_{D_N}^{+}}$ / $\bigtriangleup{G_{D_N}^{-}}$  & the insertions / deletions of nodes for $G_D$\\ \hline
$\bigtriangleup{G_{P_E}^{+}}$ / $\bigtriangleup{G_{P_E}^{-}}$& the insertions / deletions of edges for $G_P$\\ \hline
$\bigtriangleup{G_{P_N}^{+}}$ / $\bigtriangleup{G_{P_N}^{-}}$  & the insertions / deletions of nodes for $G_P$\\ \hline
$U_{Di}$/$U_{Pi}$ & one update in $\bigtriangleup{G_D}$/$\bigtriangleup{G_P}$\\ \hline
$SLen$ & \tabincell{l}{the shortest path length matrix between\\ each pair of nodes in $G_D$}\\ \hline
$\mathit{Can\_N(U_{Pi})}$ & the set of candidate nodes of $U_{Pi}$\\ \hline
$\mathit{Aff\_N(U_{Di})}$ & the set of affected nodes of $U_{Di}$\\ \hline
$AFF[u_i, v_j] = [a, b]$ & \tabincell{l}{the shortest path length from $u_i$ to $v_j$ is \\ changed from $a$ to $b$}  \\ \hline
$P_i$ & one partition \\ \hline
$IB(P_i)$/$OB(P_i)$ & the set of inner/outer bridge nodes of $P_i$\\ \hline
\end{tabular}
\label{tab:notations}
\vspace{-0.1in}
\end{table}

\subsection{ Data Graph and Pattern Graph}

\noindent \textbf{Data Graph.} A data graph is a directed graph $G_D = (V_D, E_D, f_a)$, where
\begin{itemize}
    \setlength{\itemsep}{0pt}
    \setlength{\parsep}{0pt}
    \setlength{\parskip}{0pt}
     \item $V_D$ is a set of nodes;
     \item $E_D \subseteq V_D \times V_D$, in which a tuple ($u$, $u{'}$)$\in E$ denotes a directed edge from node $u$ to $u{'}$;
     \item $f_a(u)$ is a function such that for each node $u$ $\in$ $V_D$, $f_a(u)$ is a set of labels. Intuitively, $f_a$ consists of the attributes of a node, e.g., name, age, job title \cite{lappas2011survey}.
\end{itemize}

\noindent \textbf{Example 3:} $G_D$ in Fig. \ref{fig:pnmatch}(a) depicts a data graph, where each node denotes a person, together with the label of a person, e.g., $PM$ stands for a \emph{Project Manager}. Each edge denotes a relationship between the two connected nodes, e.g., $e(PM_1, DB_1)$ means $PM_1$ has a collaboration relationship with $DB_1$.\\

\noindent \textbf{Pattern Graph.} A pattern graph is defined as $G_P = (V_P, E_P, f_v, f_e)$, where
\begin{itemize}
    \setlength{\itemsep}{0pt}
    \setlength{\parsep}{0pt}
    \setlength{\parskip}{0pt}
     \item $V_P$ and $E_P$ are a set of nodes and a set of directed edges, respectively;
     \item $f_v$ is a function defined on $V_P$ such that for each node $u \in V_P$, $f_v(u)$ is the label of node $u$, e.g., \emph{Project Manager};
     \item $f_e$ is a function defined on $E_P$ such that for each edge ($u$, $u{'}$), $f_e(u,u{'})$ is the bounded path length of $(u, u{'})$ that is either a positive integer $k$ or a symbol ``*".
\end{itemize}

\noindent \textbf{Example 4:} $G_P$ in Fig. \ref{fig:pnmatch}(b) depicts a pattern graph. In addition to the labels, each edge in $G_P$ has an integer as the bounded path length. 

\noindent \textbf{Bounded Graph Simulation (BGS).} Consider a data graph $G_D = (V_D, E_D, f_a)$ and a pattern $G_P = (V_P, E_P, f_v, f_e)$. The data graph $G_D$ \emph{matches} the pattern graph $G_P$ based on \emph{bounded graph simulation}, denoted by $G_P \unlhd G_D$, if there exists a binary relation $M \subseteq V_P \times V_D$ such that
\begin{itemize}
    \setlength{\itemsep}{0pt}
    \setlength{\parsep}{0pt}
    \setlength{\parskip}{0pt}
    \item for any $u \in V_P$, there exists $v \in V_D$, such that $(u, v) \in M$;
    \item $f_a(v)$ of $v$ includes $f_v(u)$ of $u$;
    \item for each edge $(u, u{'})$ in $E_P$, there exists a $path$ $\rho = v/.../v{'}$ in $G_D$ such that $(u{'},v{'}) \in M$, and $len(\rho) \leq k$ if $f_e(u, u{'}) = k$.
\end{itemize}

\noindent \textbf{Remark:} Note that there exists a path $\rho$ from $u$ to $u{'}$ with $len(\rho) \leq k$ if the shortest path length from $u$ to $u{'}$ is no longer than $k$. If $G_P \unlhd G_D$, the graph pattern matching results are denoted as $M(G_P, G_D)$.

\subsection{Graph Pattern based Node Matching (GPNM)}

\noindent \textbf{GPNM.} Given a pattern graph $G_P$, a data graph $G_D$, for a given node $p_i$ in $G_P$, we define the matching node of $p_i$ in $G_D$ to be $N_{p_i}$ $=$ $\{v_i|$ $v_i$ $\in$ $M(G_P, G_D)\}$, where $M(G_P, G_D)$ is the set of matching subgraphs of $G_P$ in $G_D$ based on BGS. GPNM is to find $N_{p_i}$ for $p_i$ of $G_P$ in $G_D$. If $G_D$ has no match of $G_P$ based on BGS, then $N_{p_i} = \emptyset$.

\noindent \textbf{Example 5:} Recall $G_P$ and $G_D$ shown in Fig. \ref{fig:pnmatch}(a) and Fig. \ref{fig:pnmatch}(b) respectively. Instead of finding the whole subgraphs, the GPNM aims to find the matching nodes in $G_D$ for each node of $G_P$. Taking $PM$ as an example, since $PM_1$ and $PM_2$ are in the subgraphs which can match $G_P$ based on BGS, they are the matching nodes of $PM$. The complete node matching results are shown in Table \ref{tab:GPMResult}.

\subsection{Updates-Aware Graph Pattern based Node Matching}
\begin{itemize}
    \setlength{\itemsep}{0pt}
    \setlength{\parsep}{0pt}
    \setlength{\parskip}{0pt}
    \item \textbf{Input:} a pattern graph $G_P$, a data graph $G_D$, the GPNM result of the initial query (termed as $IQuery$), a sequence of multiple updates $\bigtriangleup{G_D}$ to $G_D$, and a sequence of multiple updates $\bigtriangleup{G_P}$ to $G_P$.
    \item \textbf{Output:} the GPNM result of the subsequent query (termed as $SQuery$) of $G_{P\_new}$ in $G_{D\_new}$ ($G_{P\_new}$ and $G_{D\_new}$ denote the updated $G_P$ and $G_D$ respectively).
\end{itemize}
\noindent \textbf{Remark:} $\bigtriangleup{G_P}$ may include the insertion of edges, insertion of nodes, deletion of edges and deletion of nodes, denoted by $\bigtriangleup{G_{P_E}^{+}}$, $\bigtriangleup{G_{P_N}^{+}}$, $\bigtriangleup{G_{P_E}^{-}}$ and $\bigtriangleup{G_{P_N}^{-}}$ respectively; $\bigtriangleup{G_D}$ may include the insertion of edges, insertion of nodes, deletion of edges and deletion of nodes, denoted by $\bigtriangleup{G_{D_E}^{+}}$, $\bigtriangleup{G_{D_N}^{+}}$, $\bigtriangleup{G_{D_E}^{-}}$ and $\bigtriangleup{G_{D_N}^{-}}$ respectively. We denote each update in $\bigtriangleup{G_P}$ as $U_{Pi}$ and each update in $\bigtriangleup{G_D}$ as $U_{Di}$.

\noindent \textbf{Example 6:} Recall $G_P$ and $G_D$ shown in Fig. \ref{fig:incmatch}(c) and Fig. \ref{fig:incmatch}(a) respectively, $IQuery$ is shown in Table \ref{tab:GPMResult}. $U_{P1}$ is to insert edge $e(PM, TE)$ with a bounded path length 2 into $G_P$ and $U_{P2}$ is to insert edge $e(S, TE)$ with a bounded path length 4 into $G_P$ shown in Fig. \ref{fig:incmatch}(b); $U_{D1}$ is to insert edge $e(SE_1, TE_2)$ into $G_D$ and $U_{D2}$ is to insert edge $e(DB_1, S_1)$ into $G_D$ shown in Fig. \ref{fig:incmatch}(d). The updated pattern graph $G_{P\_new}$ and updated data graph $G_{D\_new}$ are shown in Fig. \ref{fig:incmatch}(b) and Fig. \ref{fig:incmatch}(d) respectively. The updates-aware GPNM is to deliver the $SQuery$ for the $G_{P\_new}$ in $G_{D\_new}$ based on the updates and $IQuery$.

\section{Elimination Relationships}\label{sec:Elimination Relationship}
In this section, we first analyze there types of elimination relationships. Then, we propose the effective methods to detect the elimination relationships. We further build up an index to record the hierarchical structure of these elimination relationships.
\subsection{Elimination Relationship Types}
The elimination relationships can be categorized into three types. Below we analyze the \emph{elimination relationships} for these three types respectively.\\
\noindent \textbf{Single-graph elimination relationships in $G_P$ (Type \uppercase\expandafter{\romannumeral1}):} For each update $U_{Pi}$ in pattern graph $G_P$, we need to identify the nodes in data graph $G_D$ that has the possibility to be added into or removed from the matching results. We call these nodes as \emph{candidate nodes} and put these candidate nodes into the set of candidate nodes (denoted as $Can\_N(U_{Pi})$). Given two updates $U_{Pi}$ and $U_{Pj}$, if the set of candidate nodes of an update $U_{Pi}$ covers that of $U_{Pj}$, i.e., $Can\_N(U_{Pi}) \supseteq Can\_N(U_{Pj})$, we say \emph{$U_{Pi}$ eliminates $U_{Pj}$}, denoted as $U_{Pi} \sqsupseteq U_{Pj}$.\\
\noindent \textbf{Remark:} $Can\_N(U_{Pi})$ can be divided into two subsets: a) $Can\_AN(U_{Pi})$, which represents the set of candidate nodes that has the possibility to be \emph{added into} the matching results; b) $Can\_RN(U_{Pi})$, which represents the set of candidate nodes that has the possibility to be \emph{removed from} the matching results.

\noindent \textbf{Single-graph elimination relationships in $G_D$ (Type \uppercase\expandafter{\romannumeral2}):} In GPNM, we need to investigate if the shortest path length between each pair of nodes in $G_D$ can satisfy the requirements of the bounded path length in $G_P$. For each update $U_{Di}$ in date graph $G_D$, if the shortest path between two nodes has been affected by $U_{Di}$, we call these nodes as \emph{affected nodes} and put these affected nodes into the set of affected nodes (denoted as $\mathit{Aff\_N(U_{Di})}$). Given two updates $U_{Di}$ and $U_{Dj}$, if the set of affected nodes of an update $U_{Di}$ covers that of $U_{Dj}$, i.e., $Aff\_N(U_{Dj}) \supseteq Aff\_N(U_{Di})$, we say \emph{$U_{Di}$ eliminates $U_{Dj}$}, denoted as $U_{Di} \succeq U_{Dj}$.

\noindent \textbf{Cross-graph elimination relationships between $G_P$ and $G_D$ (Type \uppercase\expandafter{\romannumeral3}):} For an update $U_{Pi}$ from a pattern graph $G_P$ and an update $U_{Di}$ from a data graph $G_D$, if these two updates keep the GPNM results unchanged, then \emph{$U_{Pi}$ and $U_{Di}$ eliminate each other}, denoted as $U_{Di} \Leftrightarrow U_{Pi}$.

\subsection{Detecting Elimination Relationships}
Below we introduce the detailed steps for detecting the three types of elimination relationships respectively.

\noindent \textbf{Detect Type \uppercase\expandafter{\romannumeral1} elimination relationships (DER-\uppercase\expandafter{\romannumeral1}):} For each update in the pattern graph, we first identify the nodes that have the possibility to be removed from or added into the original matching results. Then if the set of candidate nodes of an update $U_{Pi}$ covers that of $U_{Pj}$, then $U_{Pi}$ eliminates $U_{Pj}$. Below are the detailed steps of detecting Type \uppercase\expandafter{\romannumeral1} elimination relationships. The pseudo-code is shown in \emph{Algorithm~1}.

\setlength{\textfloatsep}{2pt}
\begin{algorithm}[t]
\tiny
\caption{DER-\uppercase\expandafter{\romannumeral1}}
\KwIn{$G_P$, $G_D$, $\bigtriangleup{G_{P}}$, $SLen$}
\KwOut{The type \uppercase\expandafter{\romannumeral1} elimination relationships of the updates}
\For{each pair of updates $U_{Pa}$ and $U_{Pb}$ $\in$ $\bigtriangleup{G_{P}}$}
{
\If{$U_{Pa}$ and $U_{Pb}$ $\in$ $\bigtriangleup{G_{P}^{-}}$}
{
\For{each pair of nodes $u_i$ and $v_i$ in $IQuery$}
{
\If{SLen($u_i$, $v_i$) $>$ the bounded path length on $U_{Pa}$}
{
Put $u_i$, $v_i$ into $\mathit{Can\_RN(U_{Pa})}$;\\
}
\If{SLen($u_i$, $v_i$) $>$ the bounded path length on $U_{Pb}$}
{
Put $u_i$, $v_i$ into $\mathit{Can\_RN(U_{Pb})}$;\\
}
}
\If{$\mathit{Can\_RN(U_{Pa})}$ $\supseteq$ $\mathit{Can\_RN}(U_{Pb})$ }
{
$U_{Pa}$ $\sqsupseteq$ $U_{Pb}$;
}
}

\If{$U_{Pa}$ and $U_{Pb}$ $\in$ $\bigtriangleup{G_{P}^{+}}$}
{
\For{each pair of nodes $u_i$ and $v_i$ in $G_D$}
{
\If{SLen($u_i$, $v_i$) $<$ the bounded path length on $U_{Pa}$}
{
Put $u_i$, $v_i$ into $\mathit{Can\_AN(U_{Pa})}$;\\
}
\If{SLen($u_i$, $v_i$) $<$ the bounded path length on $U_{Pb}$}
{
Put $u_i$, $v_i$ into $\mathit{Can\_AN(U_{Pb})}$;\\
}
}
\If{$\mathit{Can\_AN(U_{Pa})}$ $\supseteq$ $\mathit{Can\_AN}(U_{Pb})$ }
{
$U_{Pa}$ $\sqsupseteq$ $U_{Pb}$;
}
}
}
Return type \uppercase\expandafter{\romannumeral1} elimination relationships of the updates;
\end{algorithm}

\noindent \textbf{Step 1:} We first build up the shortest path length matrix, $SLen$, to record the shortest path length between each pair of nodes in $G_D$;\\
\noindent \textbf{Step 2:} For each $U_{Pi}$, if $U_{Pi}$ $\in$ $\bigtriangleup{G_{P}^{+}}$, we then inspect if the shortest path length between the pair of nodes in $IQuery$ can satisfy the bounded path length constrain on $U_{Pi}$, if not, we put these nodes into $\mathit{Can\_RN(U_{Pi})}$ as they cannot satisfy the bounded path length constrain of the newly added edge and have to be removed from $IQuery$; If $U_{Pi}$ $\in$ $\bigtriangleup{G_{D}^{-}}$, we then inspect if the shortest path length between the pair of nodes in $G_D$ can satisfy the bounded path length constrain on $U_{Pi}$, if not, we put these nodes into $\mathit{Can\_AN(U_{Pi})}$ as the edge with the shortest path length constrain they cannot satisfy has been deleted and these nodes can be added into $IQuery$;\\
\noindent \textbf{Step 3:} For each pair of $U_{Pa}$ and $U_{Pb}$ $\in$ $\bigtriangleup{G_{P}}$, if $\mathit{Can\_N(U_{Pa})}$ $\supseteq$ $\mathit{Can\_N}(U_{Pb})$, then $U_{Pa}$ $\sqsupseteq$ $U_{Pb}$.

\noindent \textbf{Remark:} In real social network-based graphs, there are many nodes having no out-degree or in-degree. Therefore, the lengths of the shortest paths from the nodes with no out-degree to other nodes, and the lengths of the shortest paths from other nodes to the nodes with no in-degree are infinite, which makes the matrix sparse. Then, we can use some techniques to compress the sparse matrix to reduce the saving space. The Hybrid format \cite{bell2009implementing} is a well-known technique that can be adopted. A storage space of size $2|N_D||K|$ is required in Hybrid format, where $|K|$ is the maximum number of non-infinite values in a row and $|N_D|$ is the number of nodes in a data graph. Compared with the space cost of $|N_D|^2$, Hybrid format can save the storage because $|K|$ is usually much smaller than $|N_D|$.

\noindent \textbf{Example 7:} Recall $G_P$ and $G_D$ shown in Fig. \ref{fig:incmatch}(c) and Fig. \ref{fig:incmatch}(a) respectively, $IQuery$ is shown in Table \ref{tab:GPMResult}. $U_{P1}$ is to insert edge $e(PM, TE)$ with a bounded path length 2 into $G_P$ and $U_{P2}$ is to insert edge $e(S, TE)$ with a bounded path length 4 into $G_P$ shown in Fig. \ref{fig:incmatch}(b). the \emph{SLen} of $G_D$ in Fig. \ref{fig:pnmatch}(c) is shown in Table \ref{tab:sdm}. With $U_{P1}$, because the $PM$ needs to be associated with $TE$ within 2 steps and the shortest path length between $PM_2$ and $TE_2$ is $\infty$, which is larger than the bounded path length 2, then $PM_2$ and $TE_2$ are added into $\mathit{Can\_RN(U_{P1})}$. After $PM_2$ and $TE_2$ are set as candidate nodes, we need to check if the nodes connected to $PM_2$ and $TE_2$ can be set as candidate nodes. Because the shortest path length between $PM_1$ and $S_1$, the shortest path length between $PM_1$ and $SE_1$, $SE_2$, and the shortest path length between$SE_2$, $SE_1$ and $TE_1$ are all less than the corresponding bounded path length in $G_P$, then only $PM_2$ and $TE_2$ are added into $\mathit{Can\_RN(U_{P1})}$; With $U_{P2}$, only $TE_2$ is added into ${Can\_RN(U_{P2})}$. The set of candidate nodes of $U_{P1}$ and $U_{P2}$ are shown in Table \ref{tab:AffnodesP}. Because ${Can\_RN(U_{P1})}$ $\supseteq$ ${Can\_RN(U_{P2})}$, then $U_{P1} \sqsupseteq U_{P2}$.
\begin{table}[h]
\tiny
\setlength{\abovecaptionskip}{0.2cm}
\setlength{\belowcaptionskip}{0cm}
\centering
\caption{$SLen$ of $G_D$ in Fig.1 (c).}
\begin{tabular}{|c|c|c|c|c|c|c|c|c|} \hline
       &$PM_1$&$PM_2$&$SE_1$&$SE_2$&$S_1$&$TE_1$&$TE_2$&$DB_1$\\ \hline
$PM_1$ &0&3&2&1&3&2&$\infty$&1\\ \hline
$PM_2$ &$\infty$&0&1&2&2&3&$\infty$&3\\ \hline
$SE_1$ &$\infty$&1&0&1&1&2&$\infty$&2 \\ \hline
$SE_2$ &$\infty$&3&2&0&3&1&$\infty$&1\\ \hline
$S_1$  &$\infty$&3&2&3&0&4&$\infty$&1\\ \hline
$TE_1$ &$\infty$&4&3&1&4&0&$\infty$&2\\ \hline
$TE_2$ &$\infty$&4&3&4&1&5&$0$&2\\ \hline
$DB_1$ &$\infty$&2&1&2&2&3&$\infty$&0\\
\hline\end{tabular}
\label{tab:sdm}
\vspace{-0.2in}
\end{table}

\begin{table}[h]
\tiny
\setlength{\abovecaptionskip}{0.2cm}
\setlength{\belowcaptionskip}{0.cm}
\centering
\caption{The set of candidate nodes of $U_{Pi}$}
\begin{tabular}{|c|c|} \hline
Updates in pattern graph&$\mathit{Can\_RN(U_{Pi})}$\\ \hline
$U_{P1}$ & $PM_2$, $TE_2$\\ \hline
$U_{P2}$ & $TE_2$\\
\hline\end{tabular}
\label{tab:AffnodesP}
\end{table}

\noindent \textbf{Theorem 1:} \emph{The order of the updates in $\bigtriangleup{G_{P}}$ does not affect the correctness of the detection of Type \uppercase\expandafter{\romannumeral1} elimination relationships}. \\
The proof of Theorem 1 can be found in the below weblink.\\ \url{http://web.science.mq.edu.au/~yanwang/Proof.pdf}. \\

\noindent \textbf{Detect Type \uppercase\expandafter{\romannumeral2} elimination relationships (DER-\uppercase\expandafter{\romannumeral2}):} For each update in the data graph, we first detect the nodes where the shortest path length in data graph between them are changed by each update (denoted as \emph{affected nodes}). Then, if the set of affected nodes of an update $U_{Di}$ covers that of $U_{Pj}$, then $U_{Pi}$ eliminates $U_{Pj}$. Below are the detailed steps of detecting Type \uppercase\expandafter{\romannumeral2} elimination relationships. The pseudo-code is shown in \emph{Algorithm 2}.

\setlength{\textfloatsep}{2pt}
\begin{algorithm}[t]
\tiny
\caption{DER-\uppercase\expandafter{\romannumeral2}}
\KwIn{$G_P$, $G_D$, $\bigtriangleup{G_{D}}$, $SLen$}
\KwOut{The type \uppercase\expandafter{\romannumeral2} elimination relationships of the updates}
\For{each pair of updates $U_{Da}$ and $U_{Db}$ $\in$ $\bigtriangleup{G_{D}}$}
{
\If{the shortest path lengths between the nodes are not affected}
{
Keep the shortest path lengths in $SLen\_new$ as that in $SLen$;\\
}
\Else
{
Apply the Dijkstra's algorithm for updating the shortest path lengths between the affected nodes in $SLen_{new}$;\\
}

Put the affected nodes into $\mathit{Aff\_N(U_{Da})}$;\\
Put the affected nodes into $\mathit{Aff\_N}(U_{Db})$;\\
\If{$\mathit{Aff\_N(U_{Da})}$ $\supseteq$ $\mathit{Aff\_N}(U_{Db})$ }
{
$U_{Da}$ $\succeq$ $U_{Db}$;
}
}
Return type \uppercase\expandafter{\romannumeral2} elimination relationships of the updates;
\end{algorithm}

\noindent \textbf{Step 1:} We first update $SLen$ to get the updated shortest path length matrix, $SLen_{new}$, for each update in data graph;\\
\noindent \textbf{Step 2:} For each update $U_{Di}$, we compare the $SLen_{new}$ with $SLen$, if the shortest path length of two nodes is changed due to $U_{Di}$, we put these nodes into $\mathit{Aff\_N}(U_{Di})$;\\
\noindent \textbf{Step 3:} For each pair of updates $U_{Da}$ and $U_{Db}$, if $\mathit{Aff\_N(U_{Da})}$ $\supseteq$ $\mathit{Aff\_N}(U_{Db})$, then $U_{Da}$ $\succeq$ $U_{Db}$.
\begin{table}[t]
\tiny
\setlength{\abovecaptionskip}{0.2cm}
\setlength{\belowcaptionskip}{-0.cm}
\centering
\caption{$SLen_{new}$ with $U_{D1}$.}
\begin{tabular}{|c|c|c|c|c|c|c|c|c|} \hline
       &$PM_1$&$PM_2$&$SE_1$&$SE_2$&$S_1$&$TE_1$&$TE_2$&$DB_1$\\ \hline
$PM_1$ &0&3&2&1&3&2&${\color{red}3}$&1\\ \hline
$PM_2$ &$\infty$&0&1&2&2&3&${\color{red}2}$&3\\ \hline
$SE_1$ &$\infty$&1&0&1&1&2&${\color{red}1}$&2 \\ \hline
$SE_2$ &$\infty$&3&2&0&3&1&${\color{red}3}$&1\\ \hline
$S_1$  &$\infty$&3&2&3&0&4&${\color{red}3}$&1\\ \hline
$TE_1$ &$\infty$&4&3&1&4&0&${\color{red}4}$&2\\ \hline
$TE_2$ &$\infty$&4&3&4&1&5&$0$&2\\ \hline
$DB_1$ &$\infty$&2&1&2&2&3&${\color{red}2}$&0\\
\hline\end{tabular}
\label{tab:sdm1}
\end{table}
\begin{table}[t]
\tiny
\setlength{\abovecaptionskip}{0.2cm}
\setlength{\belowcaptionskip}{-0.cm}
\centering
\caption{$SLen_{new}$ with $U_{D2}$.}

\begin{tabular}{|c|c|c|c|c|c|c|c|c|} \hline
       &$PM_1$&$PM_2$&$SE_1$&$SE_2$&$S_1$&$TE_1$&$TE_2$&$DB_1$\\ \hline
$PM_1$ &0&3&2&1&{\color{red}2}&2&$\infty$&1\\ \hline
$PM_2$ &$\infty$&0&1&2&2&3&$\infty$&3\\ \hline
$SE_1$ &$\infty$&1&0&1&1&2&$\infty$&2 \\ \hline
$SE_2$ &$\infty$&3&2&0&{\color{red}2}&1&$\infty$&1\\ \hline
$S_1$  &$\infty$&3&2&3&0&4&$\infty$&1\\ \hline
$TE_1$ &$\infty$&4&3&1&{\color{red}3}&0&$\infty$&2\\ \hline
$TE_2$ &$\infty$&4&3&4&1&5&$0$&2\\ \hline
$DB_1$ &$\infty$&2&1&2&{\color{red}1}&3&$\infty$&0\\
\hline\end{tabular}
\label{tab:sdm2}
\end{table}

\noindent \textbf{Remark:}When identifying the candidate nodes for the updates in pattern graph and affected nodes for the updates in data graph, we first record the shortest path length for all the pairs of nodes for once. Then for each update, we first detect the nodes where the shortest path lengths between them are unchanged; and then Dijkstra’s algorithm is applied for updating the shortest path lengths between the affected nodes.

\noindent \textbf{Example 8:} Recall $G_P$ and $G_D$ shown in Fig. \ref{fig:incmatch}(c) and Fig. \ref{fig:incmatch}(a) respectively, $IQuery$ is shown in Table \ref{tab:GPMResult}. $U_{D1}$ is to insert edge $e(SE_1, TE_2)$ into $G_D$ and $U_{D2}$ is to insert edge $e(DB_1, S_1)$ into $G_D$ shown in Fig. \ref{fig:incmatch}(d). The \emph{SLen} of $G_D$ in Fig. \ref{fig:pnmatch}(c) is shown in Table \ref{tab:sdm}. the $SLen_{new}$ of $U_{D1}$ and $U_{D2}$ in Fig. \ref{fig:pnmatch}(d) are shown in Table \ref{tab:sdm1} and Table \ref{tab:sdm2} respectively. With $U_{D1}$, because the shortest path lengths from all the other nodes to $TE_1$ are changed, then all the nodes in data graph are set as the affected nodes of $U_{D1}$. With $U_{D2}$, because the shortest path lengths from $PM_1$, $SE_2$, $TE_1$ and $DB_1$ to $S_1$ are are changed, then $PM_1$, $SE_2$, $TE_1$, $DB_1$ and $S_1$ are set as affected nodes. The set of affected nodes of $U_{D1}$ and $U_{D2}$ are shown in Table \ref{tab:AffnodesD}. Because $\mathit{Aff\_N(U_{D1})}$ $\supseteq$ $\mathit{Aff\_N}(U_{D2})$, then $U_{D1} \succeq U_{D2}$.

\begin{table}[h]
\tiny
\vspace{-0.1in}
\setlength{\abovecaptionskip}{0.2cm}
\setlength{\belowcaptionskip}{-0.cm}
\centering
\caption{The affected nodes of $U_{D1}$ and $U_{D2}$}
\begin{tabular}{|c|c|} \hline
Updates in data graph&The \emph{affected nodes}\\ \hline
$U_{D1}$ & $PM_1$, $PM_2$, $SE_1$, $SE_2$, $S_1$, $TE_1$, $TE_2$, $DB_1$\\ \hline
$U_{D2}$ & $PM_1$, $SE_2$, $S_1$, $TE_1$, $DB_1$\\
\hline\end{tabular}
\label{tab:AffnodesD}
\vspace{-0.1in}
\end{table}

\noindent \textbf{Theorem 2:} \emph{The order of the updates in $\bigtriangleup{G_{D}}$ does not affect the correctness of the detection of Type \uppercase\expandafter{\romannumeral2} elimination relationship}. \\
The proof of Theorem 2 can be found in the below weblink.\\ \emph{\url{http://web.science.mq.edu.au/~yanwang/Proof.pdf}}. \\

\noindent \textbf{Detect Type \uppercase\expandafter{\romannumeral3} elimination relationships (DER-\uppercase\expandafter{\romannumeral3}):} For an update $U_{Pi}$ from a pattern graph and an update $U_{Di}$ from a data graph, we need to inspect if these two updates keep the GPNM results unchanged. Below are the detailed steps of detecting Type \uppercase\expandafter{\romannumeral3} elimination relationships. The pseudo-code is shown in \emph{Algorithm 3}.

\setlength{\textfloatsep}{2pt}
\begin{algorithm}[t]
\tiny
\caption{DER-\uppercase\expandafter{\romannumeral3}}
\KwIn{$G_P$, $G_D$, $\bigtriangleup{G_{P}}$, $\bigtriangleup{G_{D}}$, $SLen$, $SLen_{new}$}
\KwOut{The type \uppercase\expandafter{\romannumeral3} elimination relationships  of the updates}
\For{each update $U_{Pi}$ $\in$ $\bigtriangleup{G_{P}}$}
{
Perform \emph{DER-\uppercase\expandafter{\romannumeral1}} to get $Can\_N(U_{Pi})$;
}
\For{each update $U_{Di}$ $\in$ $\bigtriangleup{G_{D}}$}
{
Perform \emph{DER-\uppercase\expandafter{\romannumeral2}} to get $Aff\_N(U_{Di})$;
}

\For{each pair of nodes $u_i$, $v_i$ in $U_{Pi}$}
{
\If{$SLen_{new}(u_i, v_i)$ $\leq$ the bounded path length on $U_{Pi}$}
{
$U_{Di}$ $\Leftrightarrow$ $U_{Pi}$;
}

}
Return type \uppercase\expandafter{\romannumeral3} elimination relationships of the updates;
\end{algorithm}
\noindent \textbf{Step 1:} For the update $U_{Pi}$ from a pattern graph, we identify the candidate nodes for $U_{Pi}$.\\
\noindent \textbf{Step 2:} For the update $U_{Di}$ from a data graph, we identify affected nodes for $U_{Di}$.\\
\noindent \textbf{Step 3:} Based on $Can\_N(U_{Pi})$ and $Aff\_N(U_{Di})$, if  $Aff\_N(U_{Di}) \supseteq Can\_N(U_{Pi})$, which means the shortest path length between any nodes in the set of candidate nodes is changed due to the update $U_{Di}$, we inspect the updated shortest path length matrix $SLen_{new}$ to check if the shortest path length of the candidate nodes can satisfy the new pattern graph. If so, no node should be added into or deleted from the matching results; therefore, $U_{Pi} \Leftrightarrow U_{Di}$.
\noindent \textbf{Example 9:} Recall $G_P$ and $G_D$ shown in Fig. \ref{fig:incmatch}(c) and Fig. \ref{fig:incmatch}(a) respectively, $IQuery$ is shown in Table \ref{tab:GPMResult}. $U_{P1}$ is to insert edge $e(PM, TE)$ with a bounded path length 2 into $G_P$ shown in Fig. \ref{fig:incmatch}(b) and $U_{D1}$ is to insert edge $e(SE_1, TE_2)$ into $G_D$ shown in Fig. \ref{fig:incmatch}(d). Based on example 7 and example 8, we have $Can\_N(U_{Pi})$=$\{PM_2, TE_2\}$ and $Aff\_N(U_{Di})$=$\{PM_1, PM_2, SE_1, SE_2, S_1, TE_1, TE_2, DB_1\}$, then $Aff\_N(U_{D1})$ $\supseteq$ $Can\_N(U_{P1})$. Since $AFF(PM_2, TE_2)$ = $(\infty, 2)$, the shortest path length between $PM_2$ and $TE_2$ can still satisfy the bounded path length on the newly inserted edge. Therefore, $U_{P1} \Leftrightarrow U_{D1}$.

\noindent \textbf{Complexity:} The complexity of the generation and the updates of $SLen$ is $\mathcal{O}(|N_D|(|N_D|+|E_D|)$ \cite{ramalingam1996computational}. In the worst case, \emph{DER-\uppercase\expandafter{\romannumeral1}}, \emph{DER-\uppercase\expandafter{\romannumeral2}} and \emph{DER-\uppercase\expandafter{\romannumeral3}} need to check $SLen_{new}$ for each update, then the complexity of each of \emph{DER-\uppercase\expandafter{\romannumeral1}}, \emph{DER-\uppercase\expandafter{\romannumeral2}} and \emph{DER-\uppercase\expandafter{\romannumeral3}} is $\mathcal{O}(|N_D|(|N_D|+|E_D|)+|\!\!\bigtriangleup\!{G}\!||N_D|^2)$, where $|N_D|$ and $|E_D|$ are the number of the nodes and the number of the edges respectively in $G_D$, and $|\!\bigtriangleup\!{G}\!|$ is the scale of the updates.

\subsection{Elimination Hierarchy Tree (EH-Tree)}
As it is computationally expensive to deliver GPNM results by investigating each of the elimination relationships among the updates, we build up an index to record the hierarchical structure of the elimination relationships. This index structure can efficiently help detect the elimination relationships between each pair of updates. We present the details of the generation of EH-Tree as follows.

\begin{figure}[h]
\vspace{-0.1in}
\centering
\includegraphics[width=0.8in,height=0.8in]{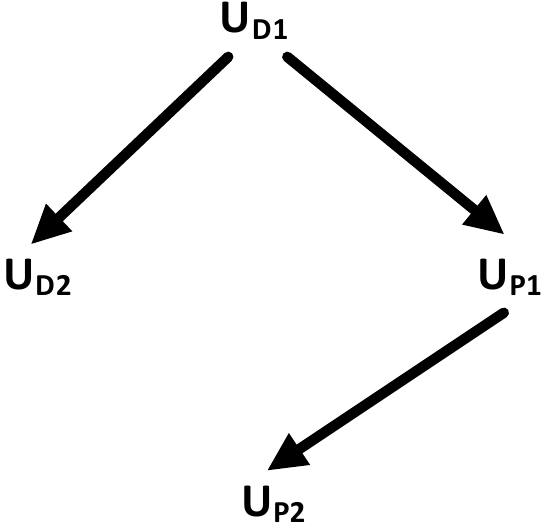}
\caption{The EH-Tree of Example 10}
\label{fig:crtree}
\end{figure}

\noindent \textbf{(1)} Firstly, for each update, we use the method mentioned in Section \ref{sec:Elimination Relationship} to identify the affected nodes for each update in data graph and identify the candidate nodes for each update in pattern graph. Each tree node in EH-Tree denotes an update and stores the affected nodes or candidates of the update.\\
\noindent \textbf{(2)} Based on the affected nodes and candidate nodes of each update, we have the following strategies: (a) the update that has the maximum number of affected nodes or candidate nodes is set as the root of an EH-Tree; (b) if the affected nodes of one update $U_{Di}$ can be covered by another update $U_{Dj}$, then $U_{Di}$ is set as a child tree node of $U_{Dj}$; (c) if the candidate nodes of one update $U_{Pi}$ can be covered by another update $U_{Pj}$, then $U_{Pi}$ is set as a child tree node of $U_{Pj}$; (d) if $U_{Di}$ and $U_{Pj}$ can eliminate each other, then we can set the $U_{Pi}$ as a child tree node of $U_{Di}$ or set the $U_{Di}$ as a child tree node of $U_{Pi}$.\\
\noindent \textbf{(3)} We then recursively insert all the updates into the EH-Tree.
\noindent \textbf{Example 10:} Recall $U_{D1}$, $U_{D2}$, $U_{P1}$ and $U_{P2}$ in Fig .\ref{fig:incmatch}. As $U_{D1}$ has the maximum number of affected nodes in all the updates, it is set as the root of EH-Tree; with $U_{D2}$, as the set of affected nodes of $U_{D1}$ covers that of $U_{D2}$, $U_{D2}$ is set as the child node of $U_{D1}$; with $U_{P1}$, as the set of candidate nodes of $U_{P1}$ covers that of $U_{P2}$, $U_{P2}$ is set as child node of $U_{P1}$; Because $U_{D1}$ and $U_{P1}$ can eliminate each other, $U_{P1}$ is set as the child node of $U_{D1}$. The completed EH-Tree is shown in Fig. \ref{fig:crtree}.
\vspace{-0.1in}

\section{Graph Partition}\label{sec:partition}

\subsection{Label-based Partition}
\begin{figure}[t]
\setlength{\abovecaptionskip}{0.cm}
\setlength{\belowcaptionskip}{-0.cm}
\centering
\includegraphics[width=2.6in,height=2.3in]{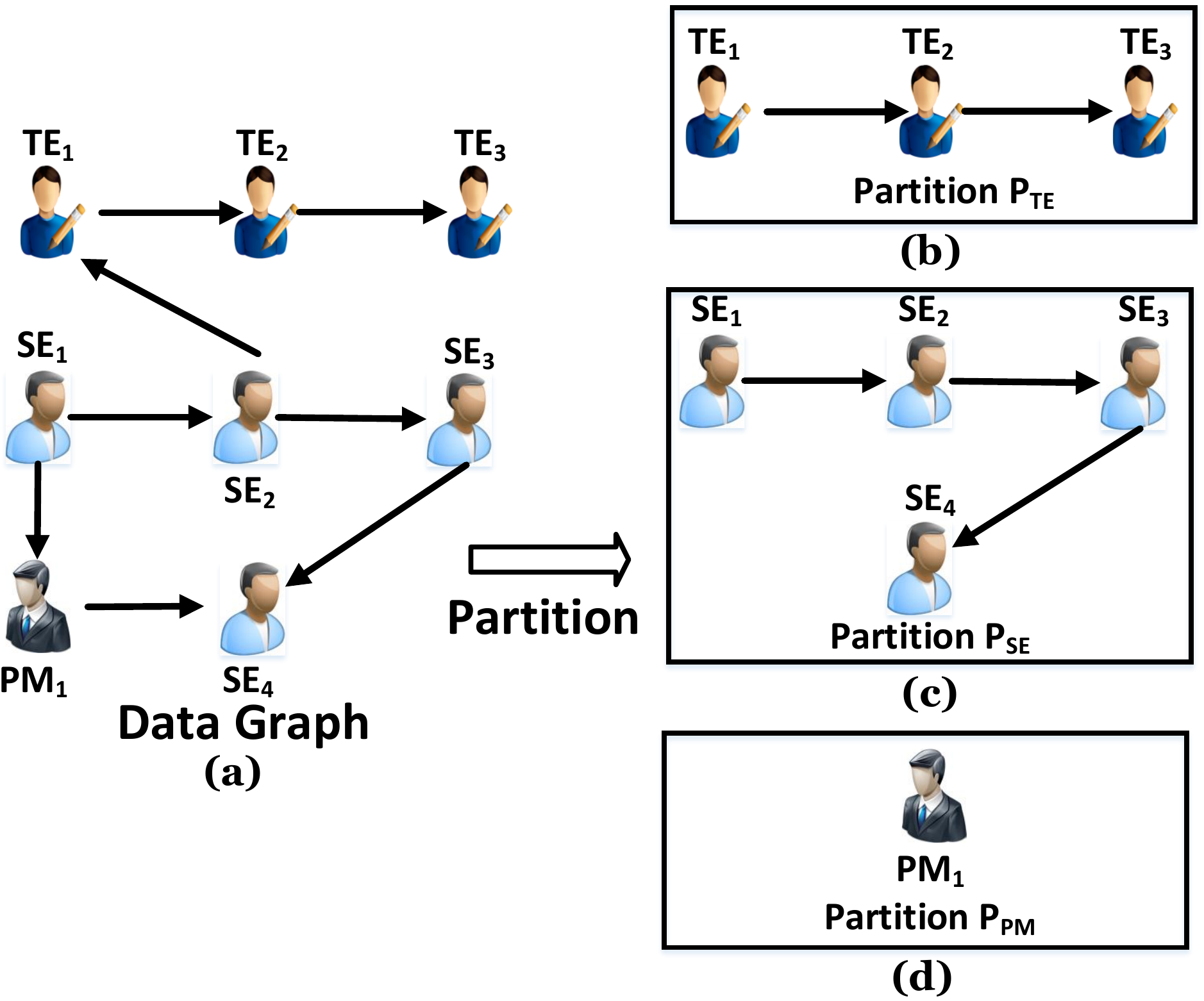}
\caption{Label-based Partition}
\label{fig:graphpartition}
\end{figure}
It is computational expensive to construct the shortest path length matrix $SLen$ and update the $SLen_{new}$. Therefore, in this section, we propose a graph partition method to improve the efficiency of computing the shortest path length between any two nodes. Based on the observation that people with the same role (e.g., has the same job title) usually connect with each other closely \cite{brandes2009structural}, we put the nodes that have the same label in a data graph and their corresponding edges into the same partition. Then the shortest path computation will be processed distributively based on the partitions.

\noindent \textbf{Example 11:} Fig. \ref{fig:graphpartition}(a) depicts a data graph, where it has three different labels of nodes, namely, $TE$, $SE$ and $PM$ respectively. Based on the different labels of the nodes, we divide the data graph into three partitions, denoted as partition $P_{TE}$, $P_{SE}$ and $P_{PM}$ respectively. 

After the partition, we need to preserve the connectivity of the data graph. Then our partition method records the cross-partition edges in the partitions where the starting nodes are in. For example, in Fig. \ref{fig:graphpartition}(a), we record $e(SE_2, TE_1)$ in the partition $P_{SE}$. Before introducing the process of computing shortest path length, we first define some nodes with properties below.
\begin{myDef}
inner bridge node: Given a partition $P_i$, a node $v_i$ ($v_i$ $\in$ $P_{i}$) is termed as an \emph{inner bridge node} of $P_i$ if there is an edge $e(v_i, v_j)$ in data graph and $v_j$ $\notin$ $P_i$. Let $IB(P_i)$ denote the set of the inner bridge nodes of partition $P_i$.
\end{myDef}

\noindent \textbf{Example 12:} In Fig. \ref{fig:graphpartition}, $SE_2$ is an inner bridge node of $P_{SE}$, because there exists an edge $e(SE_2, TE_1)$ in Fig. \ref{fig:graphpartition}(a) and $TE_1$ $\notin$ $P_{SE}$.

\begin{myDef}
outer bridge node: Given a partition $P_i$, a node $v_j$ ($v_j$ $\notin$ $P_i$) is termed as a outer bridge node of $P_i$ if there exists an edge $e(v_i, v_j)$ in data graph and $v_i$ $\in$ $P_{i}$. Let $OB(P_i)$ denote the set of the bridges nodes of partition $P_i$.
\end{myDef}

\noindent \textbf{Example 13:} In Fig. \ref{fig:graphpartition}, $PM_1$ is a outer bridge node of $P_{SE}$ because there exists an edge $e(SE_1, PM_1)$ in Fig. \ref{fig:graphpartition}(a).\\

We use record the \emph{inner bridge nodes} and \emph{outer bridge nodes} in each partition. For example, the \emph{inner bridge nodes} for partition $P_{SE}$ are $SE_1$ and $SE_2$, and the \emph{outer bridge nodes} are $PM_1$ and $TE_1$.

\subsection{Graph Partition based Shortest Path Length Computation}\label{sec:query}
We divide computation of the shortest path length into two sub-processes, i.e., \textbf{sub-process-1}: computing the shortest path length between any two nodes in the same partition, and \textbf{sub-process-2}: computing the shortest path length between any two nodes in different partitions. Below we introduce these two sub-processes in detail.

\noindent \textbf{sub-process-1:} For each partition $P_i$, if $OB(P_i)$ = $\emptyset$, we apply the Dijkstra's algorithm in this partition to compute the shortest path length. Otherwise, we apply the following steps. The pseudo-code is shown in \emph{Algorithm~4}.
\begin{itemize}
    \setlength{\itemsep}{0pt}
    \setlength{\parsep}{0pt}
    \setlength{\parskip}{0pt}
     \item \textbf{Step 1:} In each partition $P_i$, we denote the nodes in $P_i$ as $v_{Pi}$, for each pair of nodes from $v_{Pi}^a$ to $v_{Pi}^b$, we first apply the Dijkstra's algorithm to compute the shortest path length value from $v_{Pi}^a$ to $v_{Pi}^b$ in this partition (denoted as $SP_{Pi}(v_{Pi}^a, v_{Pi}^b)$) and set the shortest path length value (denotes as $SP_D(v_{Pi}^a, v_{Pi}^b)$) from $v_{Pi}^a$ to $v_{Pi}^b$ in the data graph as $SP_{Pi}(v_{Pi}^a, v_{Pi}^b)$;
     \item \textbf{Step 2:} For each outer bridge node $v_{Pj}^c$ ($v_{Pj}^c$ $\in$ $OB(P_i)$), if $OB(p_j)=\emptyset$, then the shortest path length between $v_{Pi}^a$ and $v_{Pi}^b$ is still $SP_D(v_{Pi}^a, v_{Pi}^b)$. Otherwise, if one of the outer bridge nodes of $P_j$ belongs to partition $P_i$, we combine the partitions of $P_i$ and $P_j$, and then apply the Dijkstra's algorithm to compute the shortest path length between $v_{Pi}^a$ and $v_{Pi}^b$ in the combined partition. If the new shortest path length is less than $SP_{D}(v_{Pi}^a, v_{Pi}^b)$, we update $SP_D(v_{Pi}^a, v_{Pi}^b)$ with the newly computed shortest path length;
     \item \textbf{Step 3:} We recursively apply \emph{Step 2} to update $SP_D(v_{Pi}^a, v_{Pi}^b)$ until no partition can be combined with $P_i$. 
\end{itemize}

\setlength{\textfloatsep}{2pt}
\begin{algorithm}[t]
\tiny
\caption{sub-process-1}
\KwIn{$G_D$, partitions of $G_D$}
\KwOut{The shortest path length between any two nodes in the same partition}
\For{each partition $P_i$}
{
\If{$OB(P_i)$ is $\emptyset$}
{\For{each pair of nodes from $v_{Pi}^a$ to $v_{Pi}^b$ in $P_i$}
{
Apply the Dijkstra's algorithm to compute $SP_D(v_{Pi}^a, v_{Pi}^b)$;
}
}
\Else{
\For{each pair of nodes from $v_{Pi}^a$ to $v_{Pi}^b$ in $P_i$}
{
Apply the Dijkstra's algorithm to compute $SP_{Pi}(v_{Pi}^a, v_{Pi}^b)$;\\
Set $SP_D(v_{Pi}^a, v_{Pi}^b)$=$SP_{Pi}(v_{Pi}^a, v_{Pi}^b)$;\\
\For{each outer bridge node in $P_i$ that belongs to $P_j$}
{
\If{$BN(P_j)$ is $\emptyset$}
{
Return $SP_D(v_{Pi}^a, v_{Pi}^b)$;\\
}
\Else 
{
\If{one of the outer bridge nodes in $P_j$ belongs to $P_i$}
{
Combine $P_i$ and $P_j$;\\
Apply the Dijkstra's algorithm to compute $SP_D(v_{Pi}^a, v_{Pi}^b)$ in the combined partition;\\
Update $SP_D(v_{Pi}^a, v_{Pi}^b)$;\\
}
\Else
{
Recursively inspect the outer bridge nodes of $P_j$ until no partition can be combined with $P_i$;\\
Update $SP_D(v_{Pi}^a, v_{Pi}^b)$;
}
}
}
}
}
}
Return the shortest path length between any two nodes in the same partition;
\end{algorithm}

\noindent \textbf{Example 14:} To compute the shortest path length between any two nodes in $P_{SE}$ in Fig. \ref{fig:graphpartition}, because there are two outer bridge nodes in $P_{SE}$, i.e., $PM_1$ and $TE_1$, and $P_{TE}$ has no outer bridge node and the outer bridge node of $P_{PM}$ belongs to $P_{SE}$, we combine $P_{SE}$ and $P_{PM}$. Then, we apply the Dijkstra's algorithm to compute the shortest path length in the combined partition. The shortest path length matrix of $P_{SE}$ is shown in Table \ref{tab:spse}.\\

\begin{table}[h]
\tiny
\setlength{\abovecaptionskip}{0.2cm}
\setlength{\belowcaptionskip}{-0.cm}
\centering
\vspace{-0.2in}
\caption{The shortest path length matrix of $P_{SE}$}
\begin{tabular}{|c|c|c|c|c|} \hline
       &$SE_1$&$SE_2$&$SE_3$&$SE_4$\\ \hline
$SE_1$ &0&1&2&2\\ \hline
$SE_2$ &$\infty$ &0&1&2\\ \hline
$SE_3$ &$\infty$ &$\infty$ & 0&1 \\ \hline
$SE_4$ &$\infty$ &$\infty$ & $\infty$&0 \\
\hline\end{tabular}
\label{tab:spse}
\end{table}

\noindent \textbf{sub-process-2:} For each partition $P_i$, if $OB(P_i)$ = $\emptyset$, the shortest path length from any node in $P_i$ to any node in other partitions is infinity. Otherwise, we apply the following steps. The pseudo-code is shown in \emph{Algorithm~5}.
\begin{itemize}
    \setlength{\itemsep}{0pt}
    \setlength{\parsep}{0pt}
    \setlength{\parskip}{0pt}
     \item \textbf{Step 1:} We first apply \emph{sub-process-1} to compute the shortest path length between the nodes in same partition;
     \item \textbf{Step 2:} For each inner bridge node $v_{Pi}^a$ in $P_i$ with the outer bridge node $v_{Pj}^a$, we first set $SP_D(v_{Pi}^a, v_{Pj}^a)=1$;
     \item \textbf{Step 3:} For each node $v_{Pj}^b$ in the same partition of $v_{Pj}^a$, we set $SP_D(v_{Pi}^a, v_{Pj}^b)=SP_D(v_{Pi}^a, v_{Pj}^a)+SP_D(v_{Pj}^a, v_{Pj}^b)$; And for each node $v_{Pi}^b$ in partition $P_i$,  we set $SP_D(v_{Pi}^b, v_{Pj}^b)=SP_D(v_{Pi}^b, v_{Pi}^a)+SP_D(v_{Pi}^a, vv_{Pj}^b)$. 
\end{itemize}

\setlength{\textfloatsep}{2pt}
\begin{algorithm}[t]
\tiny
\caption{sub-process-2}
\KwIn{$G_D$, partition of $G_D$}
\KwOut{The shortest path length between any two nodes in different partitions}
\For{each partition $P_i$}
{
\If{$OB(P_i)$ is $\emptyset$}
{The shortest path length from any node in $P_i$ to any node in other partitions is infinity;}
\Else
{
sub-process-1;\\
\For{each inner bridge node $v_{Pi}^a$ in $P_i$ with the outer bridge node $v_{Pj}^a$}
{
set $SP_D(v_{Pi}^a, v_{Pj}^a)=1$;
}
\For{each node $v_{Pj}^b$ in the same partition of $v_{Pj}^a$}
{
set $SP_D(v_{Pi}^a, v_{Pj}^b)=SP_D(v_{Pi}^a, v_{Pj}^a)+SP_D(v_{Pj}^a, v_{Pj}^b)$;
}
\For{each node $v_{Pi}^b$ in partition $P_i$}
{
set $SP_D(v_{Pi}^b, v_{Pj}^b)=SP_D(v_{Pi}^b, v_{Pi}^a)+SP_D(v_{Pi}^a, v_{Pj}^b)$;
}
}
}
Return the shortest path length between any two nodes in different partitions;
\end{algorithm}

\noindent \textbf{Example 15:} To compute the shortest path length from $SE$ to $TE$ in Fig. \ref{fig:graphpartition}, because $TE_1$ is the outer bridge node of $SE_2$, then we set $SP_D(SE_2, TE_1)=1$. For each node in the partition $P_{TE}$, $SP_D(SE_2, TE_2)=1+1=2$ and $SP_D(SE_2, TE_3)=1+2=3$. Since the shortest path length from $SE_3$ and $SE_4$ to $SE_2$ are infinity, the shortest path length from $SE_3$ and $SE_4$ to all the nodes in $P_{TE}$ are all infinity. The shortest path length matrix between each node in $P_{SE}$ to each node in $P_{TE}$ is shown in Table \ref{tab:spsete}.\\
\begin{table}[h]
\tiny
\setlength{\abovecaptionskip}{0.2cm}
\setlength{\belowcaptionskip}{-0.cm}
\vspace{-0.15in}
\centering
\caption{The shortest path matrix from $P_{SE}$ to $P_{TE}$}
\begin{tabular}{|c|c|c|c|} \hline
       &$TE_1$&$TE_2$&$TE_3$\\ \hline
$SE_1$ &2&3&4\\ \hline
$SE_2$ &1 &2&3\\ \hline
$SE_3$ &$\infty$ &$\infty$ & $\infty$ \\ \hline
$SE_4$ &$\infty$ &$\infty$ & $\infty$ \\
\hline\end{tabular}
\label{tab:spsete}
\vspace{-0.1in}
\end{table}

\noindent \textbf{Theorem 3:} \emph{The label-based shortest path length computation can correctly compute all-pair shortest paths}.\\
The proof of Theorem 3 can be found in the below weblink.\\ \emph{\url{http://web.science.mq.edu.au/~yanwang/Proof.pdf}}.\\

\noindent \textbf{Complexity:} In the worst case, we need to combine all the partitions to compute the shortest path length matrix. Therefore, the time complexity is $\mathcal{O}(|E_D|+|N_D|log|N_D|)$.
\section{UA-GPNM}\label{sec:UA-GPNM}
In this section, we propose a new Updates-Aware GPNM algorithm, called UA-GPNM. It first searches the EH-Tree to efficiently detect both the single-graph elimination relationships and the cross-graph elimination relationships, and then incrementally delivers the GPNM results. The detailed steps of UA-GPNM are shown below. The pseudo-code is shown in \emph{Algorithm 6}.
\setlength{\textfloatsep}{6pt}
\begin{algorithm}[t]
\setlength{\abovecaptionskip}{0.cm}
\setlength{\belowcaptionskip}{-0.cm}
\tiny
\caption{UA-GPNM}
\KwIn{$G_P$, $G_D$, $\bigtriangleup{G_{P}}$, $\bigtriangleup{G_{D}}$, $IQuery$}
\KwOut{$SQuery$}
Build up the EH-Tree for all the updates;\\
\For{each $U_{Pi}$ $\in \bigtriangleup{G_{P}}$ }
{
Check the EH-Tree;\\
\If{$U_{Pi}$ is the parent node of $U_{Pj}$ $(i\neq j)$}
{
$U_{Pi}$ can eliminate $U_{Pj}$;
}
\Else
{
\If{$U_{Pi}$ is the parent node of $U_{Di}$ $(i\neq j)$}
{
$U_{Pi}$ can eliminate $U_{Di}$;
}
}
}
\For{each $U_{Di}$ $\in \bigtriangleup{G_{D}}$ }
{
Check the EH-Tree;\\
\If{$U_{Di}$ is the parent node of $U_{Dj}$ $(i\neq j)$}
{
$U_{Di}$ can eliminate $U_{Dj}$;
}
\Else
{
\If{$U_{Di}$ is the parent node of $U_{Pi}$ $(i\neq j)$}
{
$U_{Di}$ can eliminate $U_{Pi}$;
}
}
}
Incrementally delivers the GPNM results for the updates;\\
\textbf{return} $SQuery$;
\end{algorithm}

\noindent \textbf{Step 1:} For each update $U_i$ $\in$ $\bigtriangleup{G_{D}}$ or $U_i$ $\in$ $\bigtriangleup{G_{P}}$, UA-GPNM first searches the EH-Tree to detect the elimination relationships among the updates.\\
\noindent \textbf{Step 2:} UA-GPNM then recursively finds the elimination relationships for each update until all the updates have been investigated. \\
\noindent \textbf{Step 3:} After searching the EH-Tree, we apply the incremental GPNM procedure for uneliminated updates to deliver the GPNM results. In the incremental GPNM procedure, we first find the node matching result $N_{p_i}$ of the original pattern graph $G_P$ in the original data graph $G_D$. Then, when graphs are updated, we identify the affected nodes and incrementally amend $N_{p_i}$ instead of re-computing the matching nodes from scratch. The details of the incremental GPNM procedure can be found in \cite{Sun2018incremental}.\\
\noindent \textbf{Complexity:} Since UA-GPNM first searches the EH-Tree, and then incrementally deliver the GPNM results for the updates, UA-GPNM achieves $\mathcal{O}(|N_D|(|N_D|+|E_D|)+(|\!\bigtriangleup\!{G}\!|-|U_{e}|)(|N_{D}|^{2})+|\!\bigtriangleup{G}\!|\log |\!\bigtriangleup\!{G}\!|)$ in time complexity, where $|U_e|$ is the number of the updates that can be eliminated. 

\section{Experiments}\label{sec:experiment}
\begin{figure*}
\centering
\includegraphics[height=3.1cm,width=18cm]{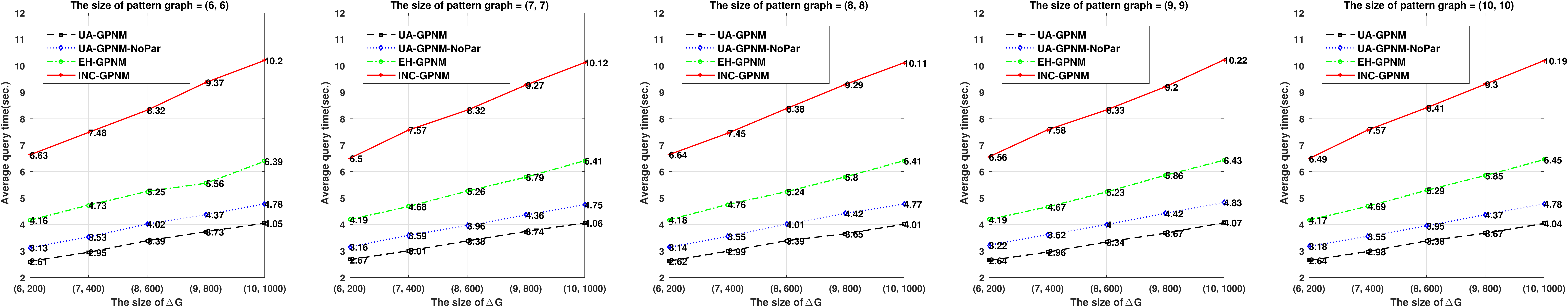}
\caption{The average query processing time in \emph{email-EU-core}}
\label{email}
\vspace{0.05in}

\centering
\includegraphics[height=3.1cm,width=18cm]{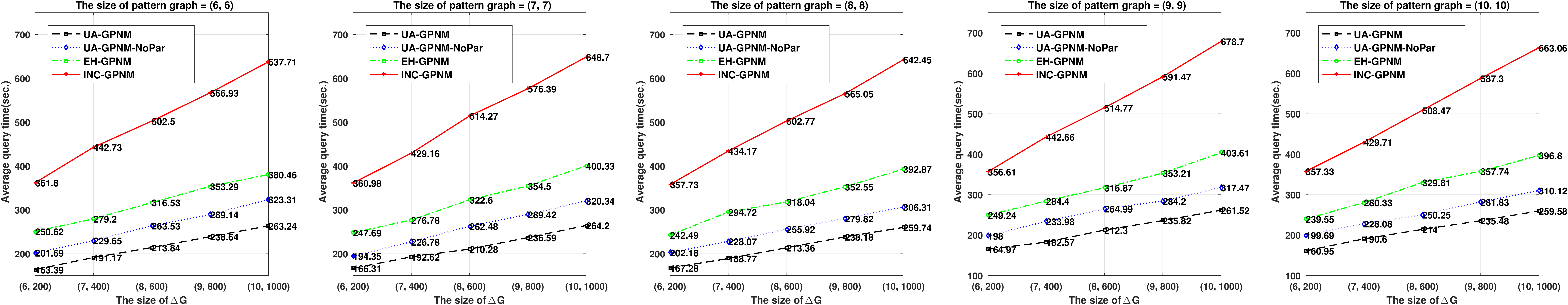}
\caption{The average query processing time in \emph{DBLP}}
\label{dblp}
\vspace{0.1in}

\centering
\includegraphics[height=3.1cm,width=18cm]{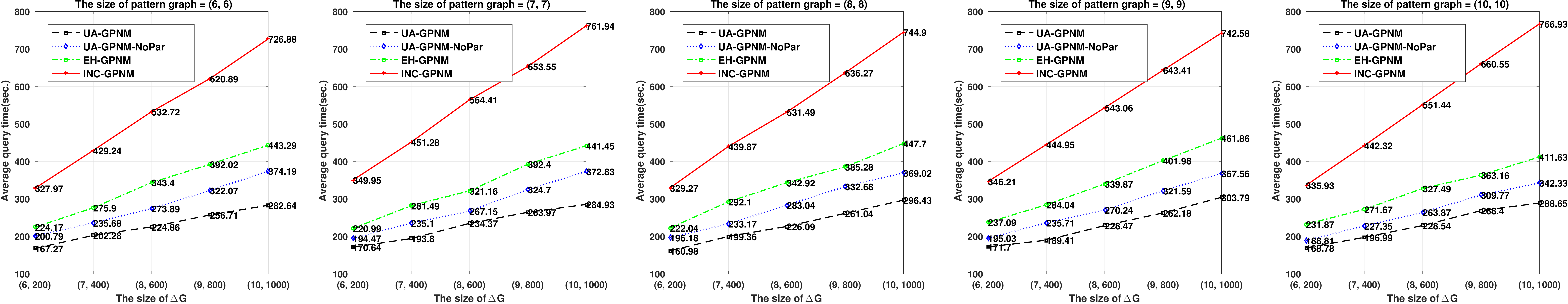}
\caption{The average query processing time in \emph{Amazon}}
\label{amazon}
\vspace{0.05in}

\centering
\includegraphics[height=3.1cm,width=18cm]{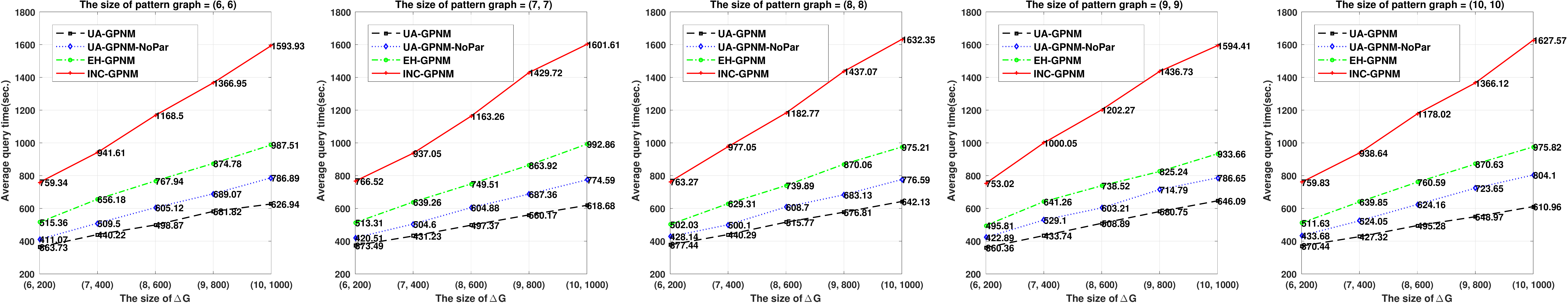}
\caption{The average query processing time in \emph{Youtube}}
\label{youtube}
\vspace{0.05in}

\centering
\includegraphics[height=3.1cm,width=18cm]{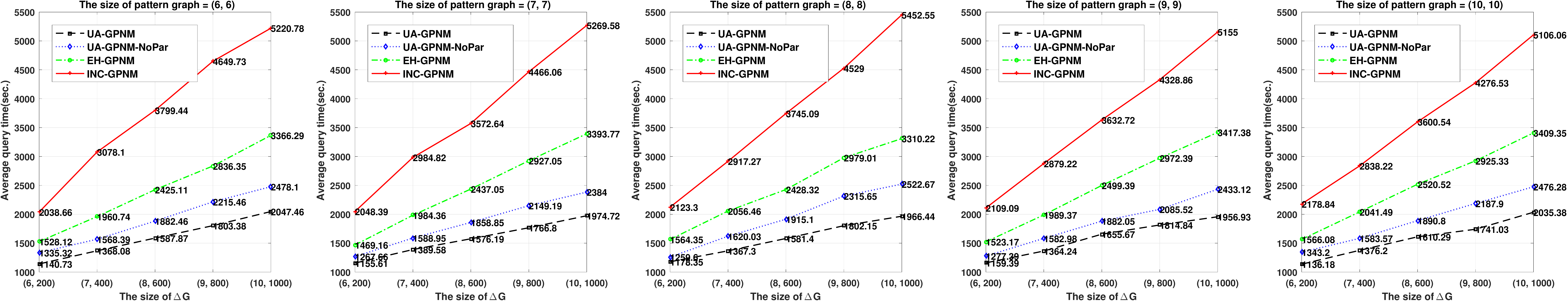}
\caption{The average query processing time in \emph{LiveJournal}}
\label{live}
\vspace{-0.2in}
\end{figure*}

We now present the results and the analysis of experiments conducted on five real-world social graphs to evaluate the performance of our proposed UA-GPNM.

\subsection{Experiment Setting}
\noindent \textbf{Datasets:} We have used five real-world social graphs that are available at \emph{snap.stanford.edu}. The details are shown in Table \ref{tab:Dataset}.

\begin{table}[h]
\tiny
\vspace{-0.2in}
\setlength{\abovecaptionskip}{0.2cm}
\setlength{\belowcaptionskip}{0cm}
\caption{The sizes of datasets}
\centering
\begin{tabular}{|c|c|c|} \hline
\textbf{Name}&\textbf{\#Nodes}&\textbf{\#Edges}\\ \hline
\emph{email-EU-core} & 1,005 &25,571 \\ \hline
\emph{DBLP} & 317,080 &1,049,866 \\ \hline
\emph{Amazon} & 334,863 &925,872 \\ \hline
\emph{Youtube} & 1,134,890 &2,987,624 \\ \hline
\emph{LiveJournal} & 3,997,962 &34,681,189 \\ \hline
\end{tabular}
\label{tab:Dataset}
\vspace{-0.1in}
\end{table}

\noindent \textbf{Pattern Graph Generation and Parameter Setting:} We used a graph generator, \emph{socnetv}\footnote{https://socnetv.org/}, to generate pattern graphs, controlled by 3 parameters: (1) the number of nodes, (2) the number of edges, and (3) the bounded path length on each edge. Since the numbers of nodes and edges in a pattern graph are usually not large \cite{fan2010graph}, they are set between 6 and 10. Since the bounded path length on each edge is usually a small integer \cite{fan2010graph}, we randomly set the bounded path length on each edge from 1 to 3.

\noindent \textbf{Updates of} \bm{$G_D$}\textbf{:} In each experiment, we removed $m_G$ edges and $m_G$ nodes from $G_D$; at the same time, we also inserted $n_G$ new edges and $n_G$ new nodes into $G_D$, where both $m_G$ and $n_G$ increase from 100 to 500 with a step of 100. \\
\noindent \textbf{Updates of} \bm{$G_P$}\textbf{:} In each experiment, we removed $m_P$ nodes and $n_P$ edges from $G_P$, and add $n_P$ new nodes and $n_P$ new edges into $G_P$, where 1 $\leq$ $m_P$ $\leq$ 5, and 1 $\leq$ $n_P$ $\leq$ 5. \\
\noindent \textbf{Remark:} In each experiment, let $\bigtriangleup{G} (\bigtriangleup{G_P}, \bigtriangleup{G_D})$ denote the updates, where $\bigtriangleup{G_P}$ denotes the updates in $G_P$ and $\bigtriangleup{G_D}$ denotes the updates in $G_D$.

\noindent \textbf{Comparison Methods:} As discussed in \emph{Section \ref{sec:relatedwork}}, there is no existing GPNM method in the literature which takes the relationships of updates and the partition strategy into consideration. Therefore, in the experiments, we implemented the following GPNM methods:
\begin{itemize}
\item \textbf{INC-GPNM}: INC-GPNM \cite{Sun2018incremental} takes the updates of $G_D$ and $G_P$ into consideration. INC-GPNM needs to perform an incremental GPNM procedure for each of the updates in $G_D$ or $G_P$.
\item \textbf{EH-GPNM}: EH-GPNM \cite{sun2019incremental} only considers the elimination relationships in data graph, when facing any update in pattern graphs, it needs to perform an incremental GPNM procedure for each of the updates in $G_P$.
\item \textbf{UA-GPNM-NoPar}: UA-GPNM-NoPar takes the relationships of updates in both pattern graph and data graph into consideration. However, it does not have the partition strategy.
\end{itemize}

\noindent \textbf{Implementation:} All the three algorithms were implemented using GCC 4.8.2 running on a server with Intel Xeon-E5 2630 2.60GHz CPU, 256GB RAM, and Red Hat 4.8.2-16 operating system. Given a data graph and a pattern graph, we consider 5 sets of updates in data graph and 5 sets of updates in pattern graph. We conduct the experiments for 5 independent runs. Therefore, in each query, there are a total of 125=5*5*5 results of the query processing time for each method and we compare the average of these results.

Figs. \ref{email}-\ref{live} depict the average query processing time with the varying sizes of $\bigtriangleup{G}$ on different sizes of $G_P$. The results and analysis are as follows.

\subsection{Experimental Results and Analysis}
\noindent $\textbf{Results-1 (Efficiency):}$ With the increase of the size of the datasets, the average processing time of UA-GPNM is always less than that of INC-GPNM, EH-GPNM and UA-GPNM-NoPar in all the cases of experiments. The detailed results are given in Table \ref{tab:15}, and the comparisons between the methods are shown in Table \ref{tab:16}. On average, (1) UA-GPNM can reduce the query processing time by \emph{58.60\%}, \emph{35.29\%} and \emph{17.70\%} compared with that of INC-GPNM, EH-GPNM and UA-GPNM-NoPar respectively. The improvement remains consistent when the size of datasets has significantly increased.
\vspace{-0.2in}
\begin{table}[h]
\tiny
\setlength{\abovecaptionskip}{0.2cm}
\setlength{\belowcaptionskip}{0.2cm}
\caption{The average query processing time based on different datasets}
\centering
\begin{tabular}{|c|c|c|c|c|} \hline
\textbf{Dataset}&\textbf{UA-GPNM}  &\textbf{UA-GPNM-NoPar} &\textbf{EH-GPNM} &\textbf{INC-GPNM} \\ \hline
\emph{email-EU-core}&3.31s& 3.98s & 5.25s & 8.27s\\ \hline
\emph{DBLP}&210.34s& 262.71s & 322.38s & 501.25s\\ \hline
\emph{Amazon}&225.48s& 278.37s & 346.15s & 536.85s\\ \hline
\emph{Youtube}&497.70s &602.41s & 753.03s & 1185.23s\\ \hline
\emph{LiveJournal}&1567.48s&1911.56s & 2449.19s & 3765.27s\\ \hline
\emph{Average}&500.86s&611.70s & 755.20s & 1199.38s\\
\hline\end{tabular}
\label{tab:15}
\vspace{-0.2in}
\end{table}

\begin{table}[h]
\tiny
\setlength{\abovecaptionskip}{0.2cm}
\setlength{\belowcaptionskip}{0.2cm}
\caption{Comparison with INC-GPNM, EH-GPNM and UA-GPNM-NoPar based on different datasets}
\centering
\begin{tabular}{|c|c|c|c|} \hline
\textbf{Dataset} &\textbf{\makecell[c]{with INC-GPNM}}&\textbf{\makecell[c]{with EH-GPNM}}&\textbf{\makecell[c]{with UA-GPNM-NoPar}} \\ \hline
\emph{email-EU-core}&\textbf{59.98\% less}& \textbf{36.95\% less}&\textbf{16.83\% less} \\ \hline
\emph{DBLP}&\textbf{58.04\% less}& \textbf{34.75\% less}&\textbf{19.77\% less} \\ \hline
\emph{Amazon}&\textbf{58.00\% less}& \textbf{34.86\% less}&\textbf{18.99\% less} \\ \hline
\emph{Youtube}&\textbf{58.60\% less}& \textbf{33.91\% less}&\textbf{14.91\% less} \\ \hline
\emph{LiveJournal}&\textbf{58.37\% less}& \textbf{36.01\% less}&\textbf{18.00\% less} \\ \hline
\emph{Average}&\textbf{58.60\% less}& \textbf{35.29\% less}&\textbf{17.70\% less} \\
\hline\end{tabular}
\label{tab:16}
\vspace{-0.1in}
\end{table}

\noindent $\textbf{Analysis-1:}$ As we discussed in \emph{Section \ref{sec:motivations}}, if there exist elimination relationships among the updates, both UA-GPNM and UA-GPNM-NoPar require less execution time than INC-GPNM and EH-GPNM as they can avoid performing an incremental GPNM procedure for each of the updates. Compared with UA-GPNM-NoPar, UA-GPNM has better efficiency as it divides the data graphs into small subgraphs, saving the query processing time when applying the the Dijkstra's algorithms.

\noindent $\textbf{Results-2 (Scalability):}$ With the increase of the scale of $\bigtriangleup{G}$ from (6, 200) to (10, 1000), the processing time of both INC-GPNM and EH-GPNM increases fast while the processing time of both UA-GPNM and UA-GPNM-NoPar increase slowly compared with that of INC-GPNM and EH-GPNM, which shows the better scalability of UA-GPNM and UA-GPNM-NoPar. Moreover, UA-GPNM has the best scalability among all the four algorithms. The detailed results are given in Table \ref{tab:18}, and the comparisons between the methods are shown in Table \ref{tab:19}.

\begin{table}[h]
\tiny
\vspace{-0.1in}
\setlength{\abovecaptionskip}{0.2cm}
\setlength{\belowcaptionskip}{0.2cm}
\caption{The average query processing time based on different scales of $\bigtriangleup{G}$}
\centering
\begin{tabular}{|c|c|c|c|c|} \hline
\makecell[c]{\textbf{Scale of} \\ \bm{$\bigtriangleup{G}$}}&\textbf{UA-GPNM}&\textbf{UA-GPNM-NoPar} &\textbf{EH-GPNM} &\textbf{INC-GPNM}\\ \hline
(6, 200) & 371.64s & 423.46s& 503.03s& 712.67s\\ \hline
(7, 400) & 439.23s & 513.71s& 643.29s& 956.63s\\ \hline
(8, 600) & 510.02s & 606.03s& 774.87s& 1182.12s\\ \hline
(9, 800) & 571.69s & 700.35s& 907.19s& 1417.40s\\ \hline
(10, 1000) & 636.42s & 786.02s& 1038.96s& 1625.27s\\
\hline\end{tabular}
\label{tab:18}
\vspace{-0.1in}
\end{table}

\begin{table}[h]
\tiny
\setlength{\abovecaptionskip}{0.2cm}
\setlength{\belowcaptionskip}{0.2cm}
\caption{Comparison with INC-GPNM, EH-GPNM and UA-GPNM-NoPar based on different scales of $\bigtriangleup{G}$}
\centering
\begin{tabular}{|c|c|c|c|} \hline
\makecell[c]{\textbf{Scale of} \\ \bm{$\bigtriangleup{G}$}}&\textbf{\makecell[c]{with INC-GPNM}}&\textbf{\makecell[c]{with EH-GPNM}}&\textbf{\makecell[c]{with UA-GPNM-NoPar}} \\ \hline
(6, 200) &\textbf{47.85\% less}&\textbf{26.12\% less}&\textbf{12.24\% less}\\ \hline
(7, 400) &\textbf{54.09\% less}&\textbf{31.72\% less}&\textbf{14.50\% less}\\ \hline
(8, 600) &\textbf{56.86\% less}&\textbf{34.18\% less}& \textbf{15.84\% less}\\ \hline
(9, 800) &\textbf{59.67\% less}&\textbf{36.98\% less}&\textbf{18.37\% less}\\ \hline
(10, 1000) &\textbf{60.84\% less}&\textbf{38.74\% less}&\textbf{19.03\% less}\\
\hline\end{tabular}
\label{tab:19}
\vspace{-0.1in}
\end{table}

\noindent $\textbf{Analysis-2:}$ With the increase of the scale of $\bigtriangleup{G}$, since INC-GPNM needs to perform an incremental GPNM procedure for each update to find the matching nodes, the scale of $\bigtriangleup{G}$ have a significant influence on their query processing time. While UA-GPNM consider the elimination relationships among the updates, the query processing time of UA-GPNM increases slowly compared with that of INC-GPNM, EH-GPNM and UA-GPNM-NoPar, which means that it has the best scalability among all the four algorithms.

\noindent $\textbf{Space Cost:}$ Since UA-GPNM uses a matrix structure to record the shortest path length between each pair of nodes and generates a balanced tree structure to index the elimination relations, its space complexity is $\mathcal{O}(|N_D|^2+|\bigtriangleup G_D||N_D|)$. Although the space cost is the same as the state-of-the-art method \cite{sun2019incremental}, our approach has much better time complexity and thus can significantly reduce query processing time (i.e., by an average of 35.29\% less than the-state-of-the-art method).

\noindent \textbf{Summary:}
The experimental results have demonstrated that the proposed UA-GPNM provides an effective means to answer GPNM queries with the updates of a data graph and a pattern graph. In addition, we have also proposed a tree structure to index the elimination relationships between the updates, and with our proposed index and partition method, UA-GPNM can greatly save query processing time. Compared to INC-GPNM, EH-GPNM and UA-GPNM-NoPar, UA-GPNM can reduce the query processing time by an average of \emph{58.60\%}, \emph{35.29\%} and \emph{17.70\%} respectively. In particular, when facing a large number of updates in a data graph, UA-GPNM has much better performance.

\section{Conclusion and future work}\label{sec:conclusion}

In this paper, we have proposed a GPNM method called UA-GPNM considering multiple updates in both data graphs and pattern graphs. UA-GPNM can efficiently deliver node matching results, and can reduce the query processing time. The experimental results on five real-world social graphs have demonstrated the efficiency of our proposed method and superiority over the state-of-art GPNM methods. In our future work, we will work on (1) the improvement on space complexity by designing new index structures, and (2) a new approach to selecting the top-k matching nodes.

\bibliographystyle{IEEEtran}
\bibliography{IEEEabrv,IEEEexample}
\end{document}